\documentclass[a4paper,fleqn,usenatbib]{mnras}   

\usepackage[T1]{fontenc}
\usepackage{ae,aecompl}
\usepackage{graphicx}	
\usepackage{amsmath}	
\usepackage{amssymb}	
\usepackage{booktabs}
\usepackage{multirow,multicol}
\usepackage{longtable,lscape}

\usepackage{rotating}
\usepackage{mathrsfs}
\usepackage{mathtools}
\usepackage{threeparttable}

\usepackage{scalefnt}

\LetLtxMacro{\oldtextsc}{\textsc}
\renewcommand{\textsc}[1]{\oldtextsc{\scalefont{1.2}#1}}
\newcommand{\cloudy}{\textsc{cloudy}}

\newcommand{\beagle}{\textsc{beagle}}

\usepackage{relsize}

\newcommand{\oi}{\hbox{[O\,{\sc i}]$\lambda6300$}}   
\newcommand{\oiid}{\hbox{[O\,{\sc ii}]$\lambda\lambda3726,3729$}}
\newcommand{\oiir}{\hbox{[O\,{\sc ii}]$\lambda3729/\lambda3726$}}
\newcommand{\oiit}{\hbox{[O\,{\sc ii}]$\lambda3727$}}
\newcommand{\oiiiaur}{\hbox{[O\,{\sc iii}]$\lambda4363$}}
\newcommand{\oiii}{\hbox{[O\,{\sc iii}]$\lambda5007$}}
\newcommand{\oiiitopt}{\hbox{[O\,{\sc iii}]$\lambda\lambda4959,5007$}}

\newcommand{\oiiid}{\hbox{O\,{\sc iii}]$\lambda\lambda1661,1666$}}
\newcommand{\oiiir}{\hbox{[O\,{\sc iii}]$\lambda4363/\lambda5007$}}
\newcommand{\oiiit}{\hbox{O\,{\sc iii}]$\lambda1663$}}
\newcommand{\oiiis}{\hbox{O\,{\sc iii}]$\lambda1666$}}
\newcommand{\nii}{\hbox{[N\,{\sc ii}]$\lambda6584$}} 
\newcommand{\siid}{\hbox{[S\,{\sc ii}]$\lambda\lambda6717,6731$}}

\newcommand{\siir}{\hbox{[S\,{\sc ii}]$\lambda6717/\lambda6731$}}

\newcommand{\nv}{\hbox{N\,{\sc v}\,$\lambda1240$}}
\newcommand{\nvd}{\hbox{N\,{\sc v}\,$\lambda\lambda1238,1242$}}
\newcommand{\civd}{\hbox{C\,{\sc iv}\,$\lambda\lambda1548,1551$}}
\newcommand{\civt}{\hbox{C\,{\sc iv}\,$\lambda1550$}}
\newcommand{\ciiid}{\hbox{[C\,{\sc iii}]$\lambda1907$+C\,{\sc iii}]$\lambda1909$}}
\newcommand{\ciiit}{\hbox{C\,{\sc iii}]$\lambda1908$}}

\newcommand{\heii}{\hbox{He\,{\sc ii}\,$\lambda1640$}}

\newcommand{\siliiid}{\hbox{[Si\,{\sc iii}]$\lambda1883$+Si\,{\sc iii}]$\lambda1892$}}
\newcommand{\siliiit}{\hbox{Si\,{\sc iii}]$\lambda1888$}}

\newcommand{\niir}{\hbox{[N\,{\sc ii}]$\lambda5755/\lambda6584$}}

\newcommand{\hii}{\hbox{H\,{\sc ii}}}

\newcommand{\ha}{\hbox{H$\alpha$}}
\newcommand{\hb}{\hbox{H$\beta$}}

\newcommand{\oiiihb}{\hbox{[O\,{\sc iii}]/H$\beta$}}
\newcommand{\oiioiii}{\hbox{[O\,{\sc ii}]/[O\,{\sc iii}]}}
\newcommand{\siiha}{\hbox{[S\,{\sc ii}]/H$\alpha$}}
\newcommand{\niiha}{\hbox{[N\,{\sc ii}]/H$\alpha$}}
\newcommand{\niioii}{\hbox{[N\,{\sc ii}]/[O\,{\sc ii}]}}

\newcommand{\ciiioiii}{\hbox{C\,{\sc iii}]/O\,{\sc iii}]}}
\newcommand{\siliiiciii}{\hbox{Si\,{\sc iii}]/C\,{\sc iii}]}}
\newcommand{\heiioiii}{\hbox{He\,{\sc ii}/O\,{\sc iii}]}}
\newcommand{\civciii}{\hbox{C\,{\sc iv}/C\,{\sc iii}]}}
\newcommand{\nvheii}{\hbox{N\,{\sc v}/He\,{\sc ii}}}
\newcommand{\ciiiheii}{\hbox{C\,{\sc iii}]/He\,{\sc ii}}}
\newcommand{\civheii}{\hbox{C\,{\sc iv}/He\,{\sc ii}}}
\newcommand{\zav}{\hbox{${Z}$}}
\newcommand{\zavsol}{\hbox{${Z}_{\odot}$}}

\newcommand{\subISM}{\textnormal{\tiny \textsc{ism}}}
\newcommand{\zism}{\hbox{$Z_\subISM$}}
\newcommand{\xid}{\hbox{$\xi_{\rm{d}}$}}
\newcommand{\xidsol}{\hbox{$\xi_{\rm{d\odot}}$}}
\newcommand{\nh}{\hbox{$n_{\mathrm{H}}$}}
\newcommand{\mup}{\hbox{$m_{\rm{up}}$}}
\newcommand{\Us}{\hbox{$U_{\rm{S}}$}}

\newcommand{\msol}{\hbox{$\mathrm{M}_{\sun}$}}

\newcommand{\nnee}{\hbox{$n_{\rm{e}}$}}
\newcommand{\Te}{\hbox{$T_{\rm{e}}$}}
\newcommand{\te}{\hbox{$t_{\rm{e}}$}}
\newcommand{\CO}{\hbox{C/O}}
\newcommand{\COsol}{\hbox{(C/O)$_\odot$}}
\newcommand{\COgas}{\hbox{(C/O)$_{\mathrm{gas}}$}}
\newcommand{\COgassol}{\hbox{(C/O)$_{\odot,\mathrm{gas}}$}}
\newcommand{\NO}{\hbox{N/O}}
\newcommand{\NOsol}{\hbox{(N/O)$_\odot$}}
\newcommand{\NOgas}{\hbox{(N/O)$_\mathrm{gas}$}}
\newcommand{\NOgassol}{\hbox{(N/O)$_{\odot,\mathrm{gas}}$}}
\newcommand{\NH}{\hbox{N/H}}

\newcommand{\OH}{\hbox{O/H}}
\newcommand{\OHgas}{\hbox{(O/H)$_\mathrm{gas}$}}

\newcommand{\OHgassol}{\hbox{(O/H)$_{\odot,\mathrm{gas}}$}}
\newcommand{\CppOpp}{\hbox{C$^{+2}$/O$^{+2}$}}
\newcommand{\comment}[1]{}

\def\apj{\mbox{ApJ}}
\def\apjl{\mbox{ApJL}}
\def\apjs{\mbox{ApJS}}
\def\aaps{\mbox{A\&AS}}
\def\mnras{\mbox{MNRAS}}
\def\aj{\mbox{AJ}}
\def\araa{\mbox{ARA\&A}}
\def\pasp{\mbox{PASP}}

\def\aap{\mbox{A\&A}}

\def\ssr{\mbox{SSR}}

\title[Nebular emission from galaxies]{Modelling the nebular emission from primeval to present-day star-forming galaxies}

\author[J. Gutkin et al.]{
Julia Gutkin$^{1}$\thanks{E-mail: gutkin@iap.fr},
St\'ephane Charlot$^{1}$ and
Gustavo Bruzual$^{2}$
\\
$^{1}$Sorbonne Universit\'es, UPMC-CNRS, UMR7095, Institut d'Astrophysique de Paris, F-75014, Paris, France\\
$^{2}$Instituto de Radioastronomía y Astrofísica, UNAM, Campus Morelia, M{\'e}xico\\
}

\date{Accepted XXX. Received YYY; in original form ZZZ}

\pubyear{2016}

\begin{document}
\label{firstpage}
\pagerange{\pageref{firstpage}--\pageref{lastpage}}
\maketitle


\begin{abstract}
We present a new model of the nebular emission from star-forming galaxies in a wide range of chemical compositions, appropriate to interpret observations of galaxies at all cosmic epochs. The model relies on the combination of state-of-the-art stellar population synthesis and photoionization codes to describe the ensemble of \hii\ regions and the diffuse gas ionized by young stars in a galaxy. A main feature of this model is the self-consistent yet versatile treatment of element abundances and depletion onto dust grains, which allows one to relate the observed nebular emission from a galaxy to both gas-phase and dust-phase metal enrichment. We show that this model can account for the rest-frame ultraviolet and optical emission-line properties of galaxies at different redshifts and find that ultraviolet emission lines are more sensitive than optical ones to parameters such as \CO\ abundance ratio, hydrogen gas density, dust-to-metal mass ratio and upper cutoff of the stellar initial mass function. We also find that, for gas-phase metallicities around solar to slightly sub-solar, widely used formulae to constrain oxygen ionic fractions and the \CO\ ratio from ultraviolet and optical emission-line luminosities are reasonable faithful. However, the recipes break down at non-solar metallicities, making them inappropriate to study chemically young galaxies. In such cases, a fully self-consistent model of the kind presented in this paper is required to interpret the observed nebular emission.
\end{abstract}


\section{Introduction}
\label{sec:intro}
The emission from interstellar gas heated by young stars in galaxies contains valuable clues about both the nature of these stars and the physical conditions in the interstellar medium (ISM). In particular, prominent optical emission lines produced by \hii\ regions, diffuse ionized gas and a potential active galactic nucleus (AGN) in a galaxy are routinely used as global diagnostics of gas metallicity and excitation, dust content, star formation rate and nuclear activity \citep[e.g.,][]{izotov99,kobulnicky99,kauffmann03,nagao06b,kewley08}. Near-infrared spectroscopy enables such studies in the optical rest frame of galaxies out to redshifts $z\sim1$--3 \citep[e.g.,][]{pettini2004,hainline09,richard11,guaita13,steidel14,shapley15}. While the future {\it James Webb Space Telescope} ({\it JWST}) will enable rest-frame optical emission-line studies out to the epoch of cosmic reionization, rapid progress is being accomplished in the observation of fainter emission lines in the rest-frame ultraviolet spectra of galaxies in this redshift range \citep[e.g.,][]{shapley03,erb10,stark14,stark15a,stark15b,stark16,sobral15}. The interpretation of these new observations requires the development of models optimised for studies of the ultraviolet -- in addition to optical -- nebular properties of chemically young galaxies, in which heavy-element abundances \citep[for example, the C/O ratio;][]{erb10,cooke11} are expected to differ substantially from those in star-forming galaxies at lower redshifts.

Several models have been proposed to compute the nebular emission of star-forming galaxies through the combination of a stellar population synthesis code with a photoionization code (e.g., \citealt{garcia95,stasinska96,charlot01}, hereafter CL01; \citealt{zackrisson01,kewley02,panuzzo03,dopita13}; see also \citealt{anders03,schaerer09}). These models have proved valuable in exploiting observations of optical emission lines to constrain the young stellar content and ISM properties of star-forming galaxies \citep[e.g.,][]{brinchmann04,blanc15}. A limitation of current models of nebular emission is that these were generally calibrated using observations of \hii\ regions and galaxies in the nearby Universe, which increasingly appear as inappropriate to study the star formation and ISM conditions of chemically young galaxies (e.g., \citealt{erb10,steidel14,steidel16,shapley15}; see also \citealt{brinchmann08,shirazi14}). We note that this limitation extends to chemical abundance estimates based on not only the so-called `strong-line' method, but also the `direct' (via the electronic temperature \Te) method, since both methods rely on the predictions of photoionization models (see Section~\ref{sec:icf}). Another notable limitation of current popular models of the nebular emission from star-forming galaxies is that these do not incorporate important advances achieved over the past decade in the theories of stellar interiors \citep[e.g.,][]{eldridge08,bressan12,ekstrom12,georgy13,chen15} and atmospheres \citep[e.g.,][]{hauschildt99,hillier99,pauldrach01,lanz03,lanz07,hamann04,martins05,puls05,rodriguez05,leitherer10}.

In this paper, we present a new model of the ultraviolet and optical nebular emission from galaxies in a wide range of chemical compositions, appropriate to model and interpret observations of star-forming galaxies at all cosmic epochs. This model is based on the combination of the latest version of the \citet{bruzual03} stellar population synthesis code (Charlot \& Bruzual, in preparation; which incorporates the stellar evolutionary tracks of \citealt{bressan12,chen15} and the ultraviolet spectral libraries of \citealt{lanz03,lanz07,hamann04,rodriguez05,leitherer10}) with the latest version of the photoionization code \cloudy\ (c13.03; described in \citealt{ferland13}). We follow CL01 and use effective (i.e. galaxy-wide) parameters to describe the ensemble of \hii\ regions and the diffuse gas ionized by successive stellar generations in a galaxy. We take special care in parametrizing the abundances of heavy elements and their depletion onto dust grains in the ISM, which allows us to model in a self-consistent way the influence of `gas-phase' and `interstellar' (i.e. gas+dust-phase) abundances on emission-line properties. We build a comprehensive grid of models spanning wide ranges of interstellar parameters, for stellar populations with a \citet{chabrier03} stellar initial mass function (IMF) with upper mass cutoffs 100 and 300\,\msol. We show that these models can reproduce available observations of star-forming galaxies in several line-ratio diagrams at optical (\oiid, \hb, \oiii, \ha, \nii, \siid) and ultraviolet (\nvd, \civd, \heii, \oiiis, \ciiid, \siliiid) wavelengths. We further exploit this model grid  to quantify the limitations affecting standard recipes based on the direct-\Te\ method to measure element abundances from emission-line luminosities. The model presented in this paper has already been used successfully to interpret observations of high-redshift star-forming galaxies \citep{stark14,stark15a,stark15b,stark16} and to define new ultraviolet and optical emission-line diagnostics of active versus inactive galaxies \citep{feltre16}.

We present our model in Section~\ref{sec:modelling}, where we parametrize the nebular emission of a star-forming galaxy in terms of stellar population, gas and dust parameters. In Section~\ref{sec:optical}, we compute a large grid of photoionization models and show that these succeed in reproducing observations of galaxies from the Sloan Digital Sky Survey (SDSS) in standard optical line-ratio diagrams. We investigate the ultraviolet properties of these models and compare them with observations of star-forming galaxies at various cosmic epochs in Section~\ref{sec:uv}. In Section~\ref{sec:icf}, we investigate the limitations of standard recipes based on the direct-\Te\ method to measure element abundances from emission-line luminosities, focusing on the \CO\ ratio as a case study. We summarise our conclusions in Section~\ref{sec:concl}.

\section[]{Modelling}
\label{sec:modelling}

To model the stellar and nebular emission from a star-forming galaxy, we adopt the isochrone synthesis technique introduced by \citet{charlot91} and express the luminosity per unit wavelength $\lambda$ emitted at time $t$ as
\begin{equation}
L_{\lambda}(t)=\int_0^t dt^\prime\, \psi(t-t^\prime) \, S_{\lambda}[t^\prime,Z(t-t^\prime)] \, T_{\lambda}(t,t^\prime)\,,
\label{eq:flux_gal}
\end{equation}
where $\psi(t-t^\prime)$ is the star formation rate at time $t-t^\prime$, $S_\lambda[t^\prime,Z(t-t^\prime)]$ the luminosity produced per unit wavelength per unit mass by a single stellar generation of age $t^\prime$ and metallicity $Z(t-t^\prime)$ and $T_\lambda(t,t^\prime)$ the transmission function of the ISM, defined as the fraction of the radiation produced at wavelength $\lambda$ at time $t$ by a generation of stars of age $t^\prime$ that is transferred by the ISM. We describe below the prescriptions we adopt for the functions $S_\lambda$ and $T_\lambda$ in equation~\eqref{eq:flux_gal}. We do not consider in this paper the potential contributions to $L_{\lambda}(t)$ by shocks nor an AGN component.

\subsection{Stellar emission}
\label{sec:stellar_code}
We compute the spectral evolution of a single stellar generation $S_\lambda[t^\prime,Z(t-t^\prime)]$ in equation~\eqref{eq:flux_gal} above using the latest version of the \citet{bruzual03} stellar population synthesis model (Charlot \& Bruzual, in preparation; see also \citealt{wofford16}). This incorporates stellar evolutionary tracks computed with the recent code of \citet{bressan12} for stars with initial masses up to 350\,\msol\ \citep{chen15} and metallicities in the range $0.0001\leq{Z}\leq0.040$ (the present-day solar metallicity corresponding to $\zavsol=0.01524$; see also Section~\ref{sec:abunddepl} below). These tracks include the evolution of the most massive stars losing their hydrogen envelope through the classical Wolf-Rayet phase (i.e., stars more massive than about $25\,\msol$ at $Z=\zavsol$, this threshold increasing as metallicity decreases). 

To compute the spectral energy distributions of stellar populations, the above evolutionary tracks are combined with different stellar spectral libraries covering different effective-temperature, luminosity-class and wavelength ranges \citep{pauldrach01,rauch02,lanz03,hamann04,martins05,rodriguez05,sanchez06,lanz07,leitherer10}. Of major interest for the present study are the prescriptions for the ionizing spectra of hot stars. For O stars hotter than 27,500\,K and B stars hotter than 15,000\,K, the ionizing spectra come from the library of metal line-blanketed, non-local thermodynamic equilibrium (non-LTE), plane-parallel, hydrostatic models of \citet{lanz03,lanz07}, which reach effective temperatures of up to 55,000\,K and cover a wide range of metallicities (from zero to twice solar). The ionizing spectra of cooler stars come from the library of line-blanketed, LTE,  plane-parallel, hydrostatic models of \citet{rodriguez05}. For O stars hotter than 55,000\,K, the ionizing spectra are taken from the library of line-blanketed, non-LTE, plane-parallel, hydrostatic models of \citet[][which are also used to describe the radiation from faint, hot post-asymptotic-giant-branch stars]{rauch02}. These are available for two metallicities, \zavsol\ and 0.1\zavsol, and interpolated in between. The lowest-metallicity spectra are used down to $\zav=0.0005$, below which pure blackbody spectra are adopted.
Finally, for Wolf-Rayet stars, the spectra come from the library of line-blanketed, non-LTE, spherically expanding models of \citet[][see also \citealt{graefener02, hamann03,hamann06,sander12,hainich14,hainich15,todt15}]{hamann04},\footnote{Available from \url{http://www.astro.physik.uni-potsdam.de/~PoWR}} available at four metallicities, $0.07\zavsol$,  $0.2\zavsol$, $0.5\zavsol$ and \zavsol. The closest of these metallicities is selected to describe the emission from Wolf-Rayet stars at any metallicity in the stellar population synthesis model.

\subsection{Transmission function of the ISM}
\label{sec:photo_code}

To compute the transmission function $T_{\lambda}(t,t^\prime)$ of the ISM in equation~\eqref{eq:flux_gal}, we follow CL01 (see also \citealt{pacifici12}) and write this as the product of the transmission functions of the ionized gas, $T_{\lambda}^{+}(t,t^\prime)$, and the neutral ISM, $T_{\lambda}^{0}(t,t^\prime)$, i.e.
\begin{equation}
T_{\lambda}(t,t^\prime)=T_{\lambda}^{+}(t,t^\prime) \, T_{\lambda}^{0}(t,t^\prime)\,.
\label{eq:trans_funct_tot}
\end{equation}
If the ionized regions are bounded by neutral material, $T_\lambda^{+} (t,t^\prime)$ will be close to zero at wavelengths blueward of the H-Lyman limit but greater than unity at the wavelengths corresponding to emission lines. In this paper, we focus on the nebular emission from star-forming galaxies, which is controlled primarily by the function $T_{\lambda}^{+}$. We assume for simplicity that this depends only on the age $t^\prime$ of the stars that produce the ionizing photons. Since 99.9 per cent of the H-ionizing photons are produced at ages less than 10\,Myr by a single stellar generation \citep[e.g.][]{charlot93, binette94}, as in CL01, we write
\begin{equation}
T_{\lambda}^{+}(t,t^\prime)=\left\{ \begin{array}{l l}
T_{\lambda}^{+}(t^\prime) \, & \mathrm{for} \hspace{3mm} t^\prime \leqslant 10\,\mathrm{Myr}\,,\\
1  &  \mathrm{for} \hspace{3mm} t^\prime > 10\,\mathrm{Myr} \,. \end{array}\right.
\label{eq:T_ionized}
\end{equation}
We do not consider in this paper the attenuation by dust in the neutral ISM, which is controlled by the function $T_{\lambda}^{0}$. We refer the reader to the `quasi-universal' prescription by \citet[][see also section~2.5 of \citealt{chevallard16}]{chevallard13} to express this quantity as a function of stellar age $t^\prime$ and galaxy inclination $\theta$, while accounting for the fact that young stars in their birth clouds are typically more attenuated than older stars in galaxies \citep[e.g.][]{silva98,charlot00}.

We use the approach proposed by CL01 to compute the transmission function of the ionized gas in equation~\eqref{eq:T_ionized}. This consists in describing the ensemble of \hii\ regions and the diffuse gas ionized by a single stellar generation in a galaxy with a set of effective parameters (which can be regarded as those of an effective \hii\ region ionized by a typical star cluster) and appealing to a standard photoionization code to compute $T_{\lambda}^{+}(t^\prime)$ at ages $t^\prime \leqslant 10\,\mathrm{Myr}$ for this stellar generation. By construction, the contributions by individual \hii\ regions and diffuse ionized gas to the total nebular emission are not distinguished in this prescription. This is justified by the fact that diffuse ionized gas, which appears to contribute around 20--50 per cent of the total H-Balmer-line emission in nearby spiral and irregular galaxies, is observed to be spatially correlated with \hii\ regions and believed to also be ionized by massive stars \citep[e.g.,][and references therein; see also \citealt{hunter90,martin97,oey97,wang97,ascasibar16}]{haffner09}. The effective parameters describing a typical \hii\ region in this approach therefore reflect the global (i.e. galaxy-wide) properties of the gas ionized by a stellar generation throughout the galaxy.

To compute $T_{\lambda}^{+}(t^\prime)$ in this context, we appeal to the latest version of the photoionization code \cloudy\ (c13.03; described in \citealt{ferland13}).\footnote{Available from \url{http://www.nublado.org}} We link this code to the spectral evolution of a single stellar generation, $S_\lambda[t^\prime,Z(t-t^\prime)]$, as achieved by CL01, to whom we refer for details. In brief, we compute the time-dependent rate of ionizing photons produced by a star cluster with effective mass $M_\ast$ as
\begin{equation}
Q(t^\prime)=\frac{M_\ast}{hc}\int_0^{\lambda_{\rm L}}{d\lambda}\,\lambda{S_\lambda}(t^\prime)\,,
\label{eq:qlyc}
\end{equation}
where $h$ and $c$ are the Planck constant and the speed of light, $\lambda_{\rm L}=912\,\AA$ is the wavelength at the Lyman limit and, for the sake of clarity, we have dropped the dependence of $S_\lambda$ on stellar metallicity \zav. In the \cloudy\ code, the gas is described as spherical concentric layers centred on the ionizing source (assumed to be pointlike). From the expression of $Q(t^\prime)$ in equation~\eqref{eq:qlyc}, we compute the time-dependent ionization parameter -- defined as the dimensionless ratio of the number density of H-ionizing photons to that of hydrogen -- at the distance $r$ from the ionizing source,
\begin{equation}
U(t^\prime,r)=Q(t^\prime)/(4\pi r^2 \nh{c})\,,
\label{eq:ur}
\end{equation}
where we have assumed for simplicity that the effective density \nh\ does not depend on the age $t^\prime$ of the ionizing stars. The dependence of $U$ on $t^\prime$ in the above expression implies that a galaxy containing several stellar generations is modelled as a mix of gas components characterised by different effective ionization parameters. As in CL01, we assume for simplicity in this paper that galaxies are ionization-bounded.

It is useful to characterise a photoionization model in terms of the ionization parameter at the Str\"omgren radius, defined by
\begin{equation}
R_{\rm S}^3(t^\prime)={3Q(t^\prime)}\slash{(4\pi n_{\rm H}^2 \epsilon \alpha_{\rm B})}\,, 
\label{eq:rs}
\end{equation}
where $\epsilon$ is the volume-filling factor of the gas (i.e., the ratio of the volume-averaged hydrogen density to \nh), assumed not to depend on $t^\prime$, and $\alpha_{\rm B}$ is the case-B hydrogen recombination coefficient \citep{osterbrock06}. The geometry of the model is set by the difference between $R_{\rm S}$ and the inner radius of the gaseous nebula, $r_{\rm in}$. For $r_{\rm in}\ll R_{\rm S}$, the thickness of the ionized shell is of the order of $R_{\rm S}$, implying spherical geometry, while for $r_{\rm in} \gtrsim R_{\rm S}$, the geometry is plane-parallel. We adopt here models with spherical geometry, in which case the ionization parameter at the Str\"omgren radius, given by (equations~\ref{eq:ur} and \ref{eq:rs})
\begin{equation}
U_{\rm S}(t^\prime) =\frac{\alpha_{\rm B}^{2/3}}{3c} \left[ \frac{3 Q(t^\prime)\epsilon^{2} n_{\rm H}}{4\pi} \right]^{1/3},
\label{eq:us}
\end{equation}
is nearly proportional to the volume-averaged ionization parameter, 
\begin{equation}
U_{\rm S}(t^\prime) \approx\langle U \rangle(t^\prime)/3\,.
\label{eq:usav}
\end{equation}
The above expression is valid for $r_{\rm in}\ll R_{\rm S}$ and neglects the weak dependence of $\alpha_{\rm B}$ on $r$ through the electron temperature.

Following CL01, we parametrize models for $T_{\lambda}^{+}(t^\prime)$ in terms of the zero-age ionization parameter at the Str\"omgren radius,\footnote{In reality, CL01 parametrized their models in terms of the zero-age, volume-averaged ionization parameter $\langle U \rangle(0)\approx3U_{\rm S}(0)$.}
\begin{equation}
U_{\rm S} \equiv U_{\rm S}(0).
\label{eq:usdef}
\end{equation}
As noted by CL01, the effective star-cluster mass $M_\ast$ in equation~\eqref{eq:qlyc} has no influence on the results other than that of imposing a maximum \Us\ at fixed \nh\ (equation~\ref{eq:ur}). In practice, we adopt values of $M_\ast$ in the range from $10^4$ to $10^7\msol$ for models with ionization parameters in the range from $\log\Us=-4.0$ to $-1.0$. The exact choice of the inner radius $r_{\rm in}$ also has a negligible influence on the predictions of models parametrized in terms of \Us, for $r_{\rm in}\ll R_{\rm S}$. As CL01, we set $r_{\rm in }\lesssim0.01\,$pc to ensure spherical geometry for all models. To evaluate $T_{\lambda}^{+}(t^\prime)$, we stop the photoionization calculations when the electron density falls below 1 per cent of \nh\ or if the temperature falls below 100\,K.

\subsection{Interstellar abundances and depletion factors}
\label{sec:abunddepl}

The chemical composition of the ISM has a primary influence on the transmission function $T_{\lambda}^{+}(t^\prime)$ of equation~\eqref{eq:T_ionized}. In this work, we adopt the same metallicity for the ISM, noted \zism, as for the ionizing stars, i.e., we set \zism=\zav. A main feature of our model is that we take special care in rigorously parametrizing the abundances of metals and their depletion onto dust grains in the ISM, to be able to model in a self-consistent way the influence of `gas-phase' and `interstellar' (i.e., total gas+dust-phase) abundances on the emission-line properties of a star-forming galaxy.

\subsubsection{Interstellar abundances}
\label{sec:abund}

The interstellar metallicity is the mass fraction of all elements heavier than helium, i.e. 
\begin{equation}
\zism =\left( \sum_{Z_i\geq3}n_iA_i\right)\Bigg/\left( \sum_{Z_i\geq1}n_iA_i\right)\,,
\label{eq:zism}
\end{equation}
where $Z_i$, $n_i$ and $A_i$ are, respectively, the atomic number, number density and atomic mass of element $i$ (with $n_1=\nh$). For all but a few species, we adopt the solar abundances of chemical elements compiled by \citet{bressan12} from the work of \cite{grevesse98}, with updates from \citet[][see table~1 of \citealt{bressan12}]{caffau11}. This corresponds to a present-day solar (photospheric) metallicity $\zavsol=0.01524$, and a protosolar (i.e. before the effects of diffusion) metallicity $Z_\odot^0=0.01774$. After some experimentation, we found that a minimal fine-tuning of the solar abundances of oxygen and nitrogen within the $\sim$1$\sigma$ uncertainties quoted by \citet{caffau11} provides a slightly better ability for the model to reproduce the observed properties of SDSS galaxies in several optical line-ratio diagrams (defined by the \oiid, \hb, \oiii, \ha, \nii\ and \siid\ emission lines; Section~\ref{sec:obs_SDSS}). Specifically, we adopt a solar nitrogen abundance 0.15~dex smaller and an oxygen abundance 0.10~dex larger than the mean values quoted in table~5 of \citet[][see also \citealt{nieva12}]{caffau11}. For consistency with our assumption $\zism=\zav$, we slightly rescale the abundances of all elements heavier than helium (by about $-0.04$~dex, as inferred using equation~\ref{eq:zism}) to keep the same total present-day solar metallicity, $\zavsol=0.01524$. Table~\ref{tab:abund_depl} lists the solar abundances adopted in this work for the elements lighter than zinc. For reference, the solar \NO\ ratio in our model is $\NOsol=0.07$.

We also wish to explore the nebular emission from star-forming galaxies with non-solar abundances. For $\zism=\zav\neq\zavsol$, we take the abundances of primary nucleosynthetic products to scale linearly with \zism\ (modulo a small rescaling factor; see below). We adopt special prescriptions for nitrogen and carbon, which are among the most abundant species:\footnote{\label{foot:cno} Carbon and nitrogen represent $\sim$26  and $\sim$6 per cent, respectively, of all heavy elements by number at solar metallicity. The most abundant element is oxygen ($\sim$48 per cent by number), often used as a global metallicity indicator.}

\begin{description}
\item \textit{\textbf{Nitrogen}}: 
abundance studies in Galactic and extragalactic \hii\ regions suggest that N has primary and secondary nucleosynthetic components \citep[e.g.,][]{garnett95,henry00b, garnett03}. Both components are thought to be synthesised primarily through the conversion of carbon and oxygen during CNO cycles in stars with masses in the range from about 4 to 8\,\msol. While the production of primary N (in CNO cycles of H burning) does not depend on the initial metallicity of the star, that of secondary N (from CO products of previous stellar generations) is expected to increase with stellar metallicity (essentially CO; see footnote~\ref{foot:cno}). Based on the analysis of several abundance datasets compiled from emission-line studies of individual \hii\ regions in the giant spiral galaxy M101 \citep{kennicutt03} and different types of starburst galaxies \citep[\hii, starburst-nucleus and ultraviolet-selected; see the compilation by][]{mouhcine02}, \citet{groves04b} find that the abundance of combined primary+secondary nitrogen can be related to that of oxygen through an expression of the type\footnote{We have introduced in equation~\eqref{eq:nitrogen} a scaling factor of 0.41 to account for the difference in solar abundances adopted here and in \citet{groves04b}.}
\begin{equation} 
\NH\approx0.41\,\OH\,\left[10^{-1.6} + 10^{(2.33 + \log\small\OH)}\right]\,,
\label{eq:nitrogen}
\end{equation}
where H, N and O correspond to $n_1$, $n_7$ and $n_8$, respectively, in the notation of equation~\eqref{eq:zism}. We implement the above prescription in our models, after which we rescale the abundances of all elements heavier than He to preserve the same \zism\ (using equation~\ref{eq:zism}). It is important to note that equation~\eqref{eq:nitrogen}, which includes our fine tuning of the solar O and N abundances, provides excellent agreement with the observational constraints on the gas-phase abundances of O and N in nearby \hii\ regions and galaxies originally used by \citet[][see references above]{groves04b}. We show this in Section~\ref{sec:depl} below (Fig.~\ref{fig:NO_OH}). 

\item \textit{\textbf{Carbon}}:  The production of C is thought to arise primarily from the triple-$\alpha$ reaction of helium in stars more massive than about 8\,\msol\ \citep[e.g.,][]{maeder92,prantzos94,gustafsson99,henry00b}, although less massive stars also produce and expel carbon \citep[e.g.,][]{vdhoek97,marigo96,marigo98,henry00a,marigo02}. Observations indicate that the \CO\ ratio correlates with the \OH\ ratio in Galactic and extragalactic \hii\ regions as well as Milky-Way stars, presumably because of the dependence of the carbon yields of massive stars on metallicity \citep[e.g.,][]{garnett95,garnett99,gustafsson99,henry00b}. Intriguingly, this trend appears to turn over at the lowest metallicities, where the \CO\ ratio increases again \citep[e.g.,][]{akerman04}. The difficulty of characterising the dependence of carbon production on metallicity has resulted in the secondary component to be either ignored (e.g., CL01, \citealt{kewley02}) or assigned a specific dependence on the \OH\ ratio \citep[e.g.,][]{dopita13} in previous models of nebular emission from \hii\ regions and galaxies. In our model, we prefer to account for the uncertainties in this component by keeping the \CO\ ratio (i.e., $n_6/n_8$ in the notation of equation~\ref{eq:zism}) as an adjustable parameter at fixed interstellar metallicity \zism\ (see Section~\ref{sec:grid} for details). Once a \CO\ ratio is adopted (in practice, by adjusting $n_6$ at fixed $n_8$), we rescale the abundances of all elements heavier than He to preserve the same \zism\ (using equation~\ref{eq:zism}). For reference, the solar \CO\ ratio in our model is $\COsol=0.44$.

\end{description}

To complete the parametrization of interstellar abundances, we must also specify that of helium. We follow \cite{bressan12} and write the He abundance by mass ($\propto n_2A_2$; equation~\ref{eq:zism}) as 
\begin{equation}
Y = Y_{\mathrm P} + (Y_{\odot}^0 - Y_{\mathrm P}) \, \zism/Z_{\odot}^0=0.2485+1.7756\,\zism\,,
\label{eq:helium1}
\end{equation}
where $Y_{\mathrm P} =0.2485$ is the primordial He abundance and $Y_{\odot}^{0}=0.28$ the protosolar one. This formula enables us to compute the helium mass fraction $Y$ at any given metallicity \zism. The hydrogen mass fraction is then simply $X=1-Y-\zism$.

\begin{table}
\begin{threeparttable}
	\centering
	\begin{tabular*}{0.45\textwidth}{r l r l}

\toprule
   Z$_i$\tnote{a} & Element & $\log(n_i/\nh)$\tnote{b} & $(1-f_{\rm dpl}^i$)\tnote{c}\\
\midrule
  2 & He & $-1.01$ & 1   \\
  3 & Li & $-10.99$ & 0.16\\
  4 & Be & $-10.63$ & 0.6 \\
  5 & B & $-9.47$ & 0.13 \\
  6 & C & $-3.53$ & 0.5 \\
  7 & N & $-4.32$ & 1 \\
  8 & O & $-3.17$ & 0.7 \\
  9 & F & $-7.47$ & 0.3 \\
  10 & Ne & $-4.01$ & 1 \\
  11 & Na & $-5.70$ & 0.25 \\
  12 &  Mg & $-4.45$ & 0.2 \\
  13 &  Al & $-5.56$ & 0.02 \\
  14 & Si & $-4.48$ & 0.1 \\
  15 & P & $-6.57$ & 0.25 \\
  16 & S & $-4.87$ & 1 \\
  17 & Cl & $-6.53$ & 0.5 \\
  18 & Ar & $-5.63$ & 1 \\
  19 & K & $-6.92$ & 0.3 \\
  20 & Ca & $-5.67$ & 0.003 \\
  21 & Sc & $-8.86$ & 0.005 \\
  22 & Ti & $-7.01$ & 0.008 \\
  23 & V & $-8.03$ & 0.006 \\
  24 & Cr & $-6.36$ & 0.006 \\
  25 & Mn & $-6.64$ & 0.05 \\
  26 & Fe & $-4.51$ & 0.01 \\
  27 & Co & $-7.11$ & 0.01 \\
  28 & Ni & $-5.78$ & 0.04 \\
  29 & Cu & $-7.82$ & 0.1 \\
  30 & Zn & $-7.43$ & 0.25 \\
\bottomrule
\end{tabular*}
\begin{tablenotes}
\item [a] Atomic number
\item [b] Abundance by number relative to hydrogen
\item [c] $f_{\rm dpl}^i$ is the fraction of element $i$ depleted onto dust grains (the non-refractory elements He, N, Ne, S and Ar have $f_{\rm dpl}^i=0$)
\end{tablenotes}
\caption{Interstellar abundances and depletion factors of the 30 lightest chemical elements for $\zism=\zavsol=0.01524$ and $\xid=\xidsol=0.36$ (see text for details).}
\label{tab:abund_depl}
\end{threeparttable}
\end{table}

\subsubsection{Depletion factors}
\label{sec:depl}

In our model, we account for the depletion of refractory metals onto dust grains. Observational determinations of depletion factors in Galactic interstellar clouds show a large dispersion depending on local conditions \citep[e.g.,][]{savage96}. For simplicity, we adopt the default ISM depletion factors of \cloudy\ for most elements, with updates from \citet[][see their table~1]{groves04b} for C, Na, Al, Si, Cl, Ca and Ni. By analogy with our fine-tuning of the N and O abundances in Section~\ref{sec:abund}, we slightly adjust the fraction of oxygen depleted from the gas phase (from 40 to 30 per cent for $\xid=\xidsol$) to improve the model agreement with  observed properties of SDSS galaxies in several optical line-ratio diagrams (Section~\ref{sec:grid}). The depletion factors adopted in this work are listed in Table~\ref{tab:abund_depl} for the elements lighter than zinc.

The elements depleted from the gas phase make up the grains, for which in \cloudy\ we adopt a standard \citet{MRN} size distribution and optical properties from \cite{martin91}. These grains influence radiative transfer via absorption and scattering of the incident radiation, radiation pressure, collisional cooling and photoelectric heating of the gas \citep[see, e.g.,][for a description of the influence of these effects on nebular emission]{shields95,dopita02,groves04a}. In addition, the depletion of important coolants from the gas phase reduces the efficiency of gas cooling through infrared fine-structure transitions, which causes the electron temperature to rise, thereby increasing cooling through the more energetic optical transitions. 

Following CL01, we explore the influence of metal depletion on the nebular emission from star-forming galaxies by means of the dust-to-metal mass ratio parameter, noted \xid. For $\zism=\zavsol$, the values in Table~\ref{tab:abund_depl} imply that 36 per cent by mass of all heavy elements are in the solid phase, i.e., $\xidsol=0.36$. To compute the depletion factor $f_{\rm dpl}^i$ of a given refractory element $i$ for other dust-to-metal mass ratios in the range $0\leq\xid\leq1$, we note that this must satisfy $f_{\rm dpl}^i=0$ and $1$ for $\xid=0$ and 1, respectively. We use these boundary conditions and the data in Table~\ref{tab:abund_depl} to interpolate linearly $f_{\rm dpl}^i$ as a function of $\xid$ for $\xid\neq\xidsol$.

\begin{figure}\includegraphics[width=\columnwidth]{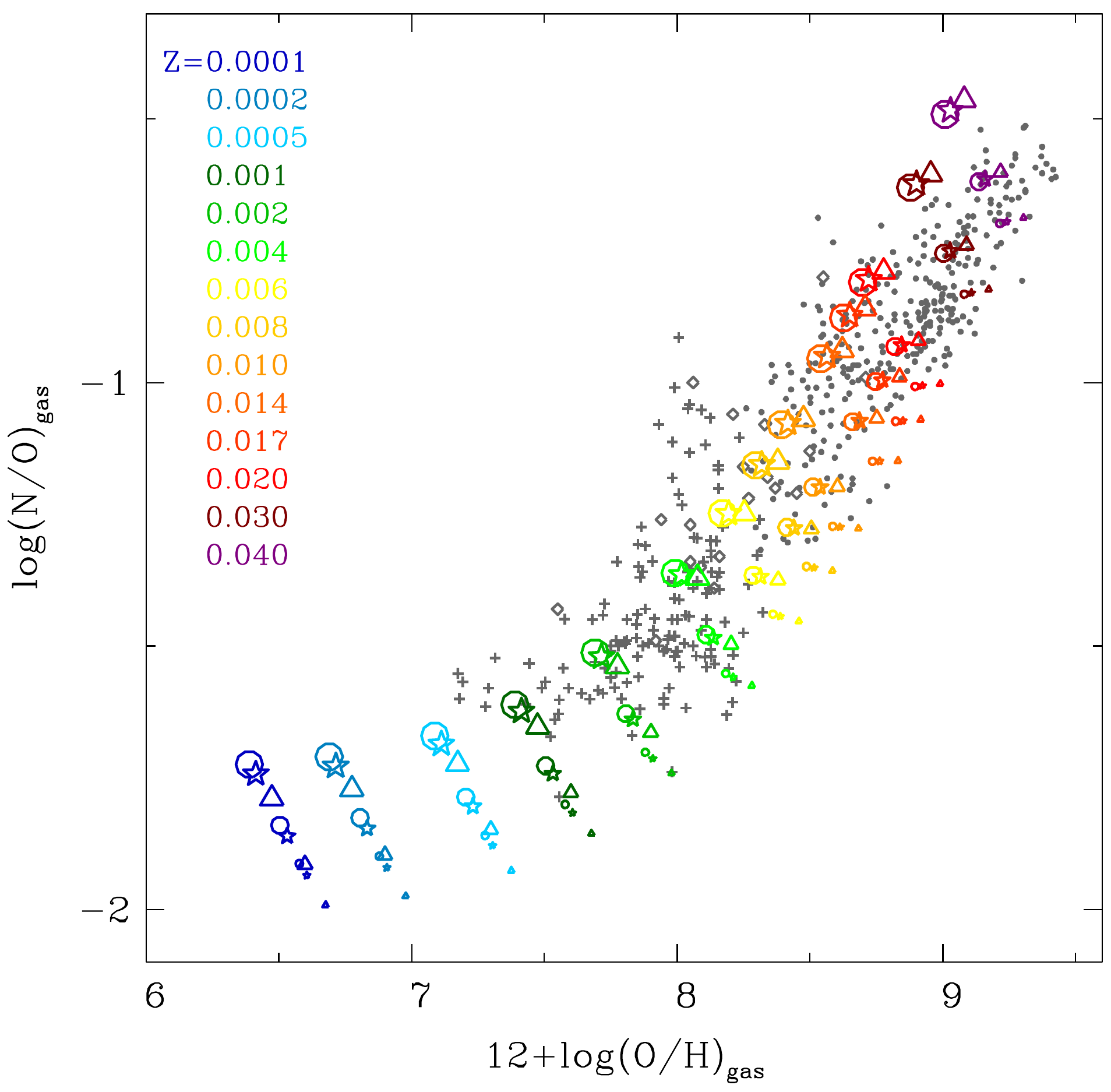}
\caption{$\log\NOgas$ as a function of $12+\log\OHgas$, for models with interstellar metallicity $\zism=0.0001$, 0.0002, 0.0005, 0.001, 0.002, 0.004, 0.006, 0.008, 0.010, 0.014, 0.017, 0.020, 0.030 and 0.040 (colour-coded as indicated), carbon-to-oxygen abundance ratio $\CO=0.1$ (triangle), 1.0 (star) and 1.4 (circle) times \COsol, and dust-to-metal mass ratio $\xid=0.1$, 0.3 and 0.5 (in order of increasing symbol size). The data (identical to those in fig.~2 of \citealt{groves04b}) are abundance datasets compiled from emission-line studies of individual \hii\ regions in the giant spiral galaxy M101 \citep[][open diamonds]{kennicutt03} and different types of starburst galaxies (\hii\ [crosses] and starburst-nucleus [dots]; see also \citealt{mouhcine02}).}
\label{fig:NO_OH}
\end{figure}

It is instructive to examine the differences in gas-phase oxygen abundance, \OHgas, corresponding to different plausible values of \xid\ at fixed interstellar metallicity \zism\ in this context. Table~\ref{tab:12logOH} lists \OHgas\ for the 14 metallicities at which stellar population models are available in the range $0.0001\leq{Z}\leq0.040$ (and for the present-day solar metallicity \zavsol=0.01524; Section~\ref{sec:stellar_code}), for 3 dust-to-metal mass ratios, $\xid=0.1$, 0.3 and 0.5, and fixed \COsol\ ratio. The gas-phase oxygen abundance can change by typically 0.2\,dex depending on the adopted $\xid$, the difference with the interstellar (gas+dust-phase) abundance, \OH, reaching up to 0.25\,dex for $\zism=\zavsol$, for example. Fig.~\ref{fig:NO_OH} shows \NOgas\ as a function \OHgas\ for the 14 metallicities and 3 dust-to-metal mass ratios in Table~\ref{tab:12logOH}, and for 3 values of the \CO\ ratio, 0.1, 1.0 and 1.4 times \COsol. The models compare well with observational constraints on these quantities in nearby \hii\ regions and star-forming galaxies, also shown on the figure (the data in Fig.~\ref{fig:NO_OH} are the same as those in fig.~2 of \citealt{groves04b}). For reference, for solar values of the metallicity, \zavsol, dust-to-metal mass ratio, \xidsol, and \CO\ ratio, \COsol, the gas-phase abundances are $12+\log\OHgassol=8.68$,  $\NOgassol=0.10$ and $\COgassol=0.31$. The large spread in \NOgas\ and \OHgas\ for models with different \xid\ and \CO\ at fixed metallicity in Fig.~\ref{fig:NO_OH} emphasizes the importance of distinguishing gas-phase from interstellar metallicity when studying the metal content of star-forming galaxies.

\begin{table}
\begin{threeparttable}
\centering 
\begin{tabular*}{0.45\textwidth}{l c c c c}
\toprule
   \zism\ & $12+\log\OH$ & \multicolumn{3}{c}{$12+\log\OHgas$} \\
\cmidrule{3-5}  
  &   & \xid=0.1 & \xid=0.3  & \xid=0.5 \\
\midrule
  0.0001 & 6.64  & 6.61  &  6.53   & 6.41   \\
  0.0002 & 6.94 & 6.91   & 6.83   & 6.71  \\
  0.0005 & 7.34 & 7.30   & 7.23   & 7.11   \\
  0.001 & 7.64 & 7.61   & 7.53   & 7.41    \\
  0.002 & 7.94 & 7.91   & 7.83   & 7.71  \\
  0.004 & 8.24 & 8.21  & 8.14   & 8.02    \\
  0.006 & 8.42 & 8.39  & 8.31   & 8.19  \\
  0.008 & 8.55 & 8.52  & 8.44   & 8.32   \\
  0.010 & 8.65  & 8.61  & 8.54   & 8.42  \\
  0.014 & 8.80  & 8.76  & 8.69   & 8.56  \\
  0.01524 (\zavsol) & 8.83 & 8.80  & 8.71   & 8.58   \\
  0.017 & 8.88  & 8.85  & 8.77   & 8.65  \\
  0.020 & 8.96  & 8.92  & 8.85   & 8.72  \\
  0.030 & 9.14  & 9.11  & 9.03  & 8.90   \\
  0.040 & 9.28 & 9.24 & 9.16  & 9.03 \\
\bottomrule
\end{tabular*}
\caption{Oxygen abundances for interstellar metallicities \zism\ corresponding to the 14 metallicities at which stellar population models are available in the range $0.0001\leq{Z}\leq0.040$ (and for the present-day solar metallicity \zavsol=0.01524), assuming $\CO=\COsol$. The second column lists the total interstellar (i.e. gas+dust-phase) oxygen abundance, $12+\log\OH$, while the three rightmost columns indicate the gas-phase oxygen abundances, $12+\log\OHgas$, corresponding to three choices of the dust-to-metal mass ratio, $\xid=0.1$, 0.3 and 0.5.}
\label{tab:12logOH}
\end{threeparttable}
\end{table}

\section{Optical emission-line properties}
\label{sec:optical}
In this section, we build a comprehensive grid of photoionization models of star-forming galaxies using the approach described in Section~\ref{sec:modelling}. We show how these models succeed in reproducing the optical emission-line properties of observed SDSS galaxies and explore the influence of the various adjustable model parameters on predicted line-luminosity ratios. 

\subsection{Grid of photoionization models}
\label{sec:grid}

Our motivation is to build a grid of photoionization models that should be adequate for the purpose of investigating the emission-line properties of star-forming galaxies at all cosmic epochs. To this end, we adopt the following sampling of the main adjustable model parameters described in Section~\ref{sec:modelling}. We emphasize that these must be regarded as effective parameters describing the ensemble of \hii\ regions and the diffused gas ionized by young stars in a galaxy:

\begin{description}
\item {\textit{\textbf{Interstellar metallicity, $\boldsymbol{Z_\subISM}$}}}:  we consider 14 values of \zism\ between 0.6 per cent of and 2.6 times solar, corresponding to metallicities at which stellar population models are available (Section~\ref{sec:stellar_code}). These are $\zism=0.0001$, 0.0002, 0.0005, 0.001, 0.002, 0.004, 0.006, 0.008, 0.010, 0.014, 0.017, 0.020, 0.030 and 0.040.

\item {\textit{\textbf{Zero-age ionization parameter at the Str\"omgren radius, $\boldsymbol{U_{\rm{S}}}$}}}: we compute models for 7 values of \Us\ logarithmically spaced in the range $-4\leq\log\Us\leq-1$, in bins of 0.5~dex.

\item {\textit{\textbf{Dust-to-metal mass ratio $\boldsymbol{\xid}$}}}: we adopt 3 choices for the dust-to-metal mass ratio, $\xid=0.1$, 0.3 and 0.5 (\xidsol=0.36).

\item {\textit{\textbf{Carbon-to-oxygen abundance ratio, $\boldsymbol{\CO}$}}}: we consider 9 logarithmically spaced values of the interstellar \CO\ ratio, corresponding to factors of 0.10, 0.14, 0.20, 0.27, 0.38, 0.52, 0.72, 1.00 and 1.40 times the solar value, \COsol=0.44.

\item {\textit{\textbf{Hydrogen gas density, $\boldsymbol{\nh}$}}}: we compute models for 4 hydrogen densities of the ionized gas, $\nh=10$, $10^2$, $10^3$ and $10^4{\rm cm}^{-3}$. These span most of the range of observed electronic densities in extragalactic \hii\ regions \citep[e.g.,][]{hunt09}.

\item {\textit{\textbf{Upper mass cutoff of the IMF $\boldsymbol{\mup}$}}}: we adopt the Galactic-disc IMF of \citet{chabrier03}, for which we fix the lower mass cutoff at $m_{\mathrm L}=0.1\,\msol$ and consider 2 values of the upper mass cutoff, $\mup=100$ and 300\,\msol.
\end{description}

Table~\ref{tab:parameters} summarises the above sampling of the main adjustable model parameters, which leads to a total of 21,168 photoionization models, each computed at 90 stellar population ages $t^\prime$ between 0 and 10\,Myr (equation~\ref{eq:T_ionized}). In the remainder of this paper, we compute all predictions of nebular emission from star-forming galaxies assuming star formation at a constant rate, $\psi(t-t^\prime)=1\,\msol{\rm yr}^{-1}$ (equation~\ref{eq:flux_gal}), for 100\,Myr. Since most ionizing photons are released during the first 10\,Myr of evolution of a single stellar generation (Section~\ref{sec:photo_code}), assuming a constant star formation rate for 100\,Myr ensures that a steady population of \hii\ regions is established in the model galaxy. In this context, the age of 100\,Myr should be interpreted as the effective age of the most recent episode of star formation in the galaxy. Finally, we assume for simplicity that all stars in a given galaxy have the same metallicity, i.e., we write 
\begin{equation}
S_{\lambda}[t^\prime,\zav(t-t^\prime)]\equiv S_{\lambda}(t^\prime,\zav)
\label{eq:zsame}
\end{equation} 
in equation~\eqref{eq:flux_gal}.

\begin{table}
\begin{threeparttable}
\centering 
\begin{tabular*}{0.47\textwidth}{l l}
\toprule
Parameter & \ \ Sampled values\\
\midrule
\zism & \ \ 0.0001, 0.0002, 0.0005, 0.001, 0.002, 0.004, 0.006, \\
         & \ \ 0.008, 0.010, 0.014, 0.017, 0.020, 0.030, 0.040\\
$\log\Us$ & \ \ $-1.0, -1.5, -2.0, -2.5, -3.0, -3.5, -4.0$\\
\xid & \ \ 0.1, 0.3, 0.5\\
$\log(\nh/{\rm cm}^{-3})$ & \ \ 1, 2, 3, 4\\
(\CO)/\COsol\ & \ \ 0.10, 0.14, 0.20, 0.27, 0.38, 0.52, 0.72, 1.00, 1.40\\
$\mup/\msol$ & \ \ 100, 300\\
\bottomrule
\end{tabular*}
\caption{Grid sampling of the main adjustable parameters of the photoionization model of star-forming galaxies described in Section~\ref{sec:modelling} (see Section~\ref{sec:grid}  for details).}
\label{tab:parameters}
\end{threeparttable}
\end{table}

\subsection{Comparison with observations}
\label{sec:obs_SDSS}

\begin{figure*}
\begin{center}
  \includegraphics[scale=0.75]{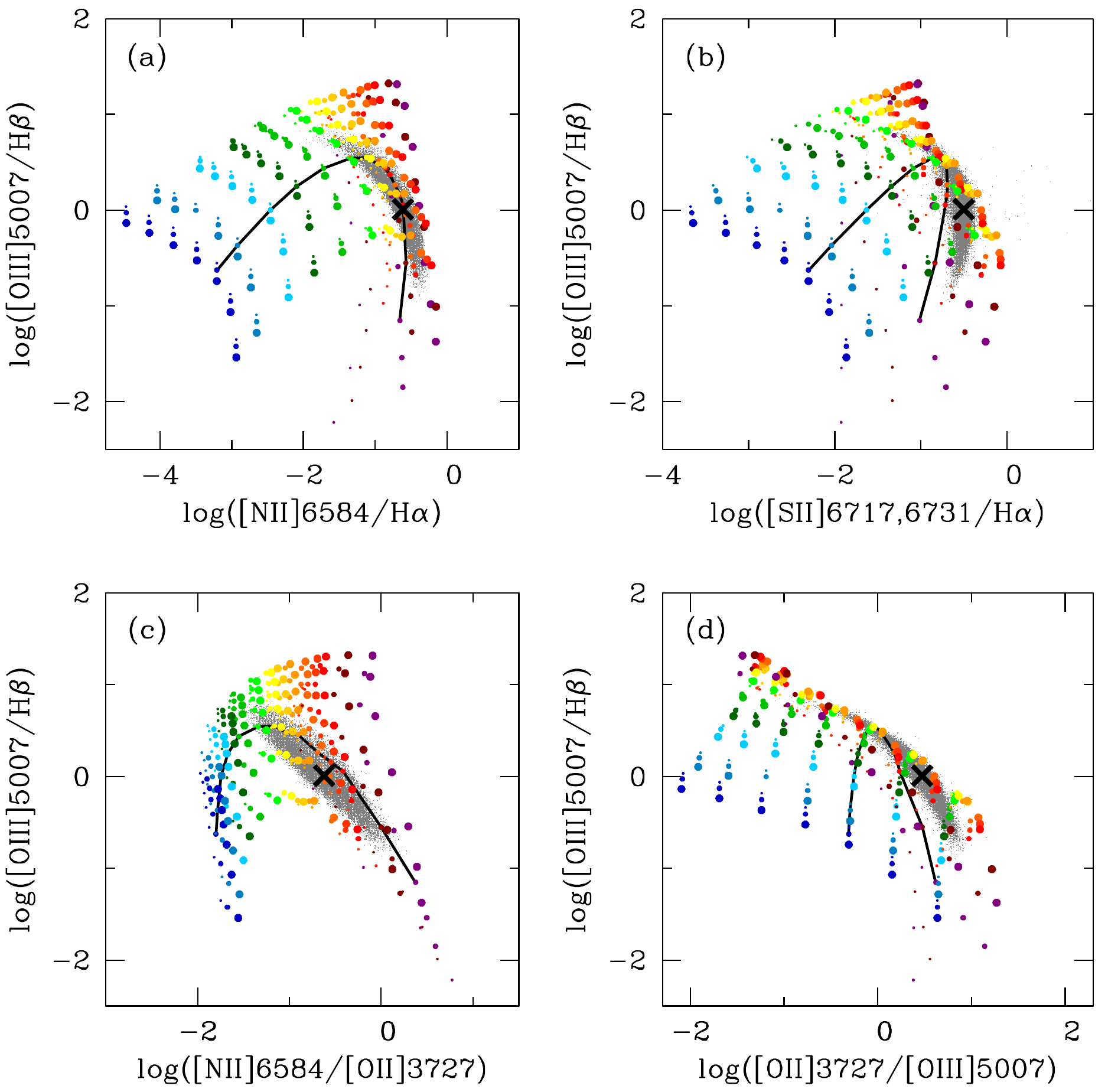}
  \caption[models]{Luminosity ratios of prominent optical emission lines predicted by the photoionization models  described in Section~\ref{sec:modelling}: (a) \oiii/\hb\  against \niiha; (b) \oiii/\hb\ against \siid/\ha; (c) \oiii/\hb\ against \nii/\oiit; and (d) \oiii/\hb\ against \oiit/\oiii.  The models assume constant star formation over the past 100\,Myr and fixed hydrogen density, $\nh=100\,{\rm cm}^{-3}$, carbon-to-oxygen ratio, \COsol=0.44, and IMF upper mass cutoff, $\mup=100\,\msol$.  They are shown for 14 interstellar metallicities in the range $0.0001\leq\zism\leq0.040$ (colour-coded as in Fig.~\ref{fig:NO_OH}; see Table~\ref{tab:12logOH}), 7 zero-age ionization parameters in the range $-4\leq\log\Us\leq-1$, in bins of 0.5~dex (in order of increasing \oiiihb\ ratio at fixed metallicity), and 3 dust-to-metal mass ratios $\xid=0.1$, 0.3 and 0.5 (in order of increasing symbol size). In each panel, a line links models with $\log\Us=-3.0$ at all metallicities, while the black cross shows the `standard' model defined in Section~\ref{sec:obs_SDSS}. The grey dots show high-quality observations of star-forming galaxies from the SDSS DR7, corrected for attenuation by dust as described in \citet{brinchmann04}.}
  \label{fig:entire_grid}
\end{center}
\end{figure*}

In Fig.~\ref{fig:entire_grid}, we compare the predictions of a subset of the grid of photoionization models described in Section~\ref{sec:grid} with high-quality observations of emission-line ratios in the spectra of 28,075 low-redshift ($0.04 \le z \le 0.2$) star-forming galaxies from the Sloan Digital Sky Survey Data Release 7 \citep[SDSS DR7][]{abazajian09}. The models span full ranges of metallicity, \zism, ionization parameter, \Us, and dust-to-metal mass ratio, \xid, for fixed gas density, $\nh=100\,{\rm cm}^{-3}$, carbon-to-oxygen ratio, $\COsol=0.44$, and IMF upper mass cutoff, $\mup=100\,\msol$. The observational sample was assembled by \cite{pacifici12}, who selected SDSS-DR7 galaxies with signal-to-noise ratio greater than 10 in all optical emission lines used to construct the data plotted in Fig.~\ref{fig:entire_grid}, i.e., \oiid\ (hereafter \oiit), \hb, \oiii, \ha, \nii\ and \siid. In doing so, \citet{pacifici12} excluded galaxies for which nebular emission could be contaminated by an AGN, according to the conservative criterion of \citealt{kauffmann03} in the standard line-diagnostic diagram of \citet[][see also \citealt{veilleux87}]{baldwin81} defined by the \oiii/\hb\ and \nii/\ha\ ratios (Fig.~\ref{fig:entire_grid}a).

Fig.~\ref{fig:entire_grid} shows that the model grid considered here succeeds in accounting for the observed properties of SDSS star-forming galaxies. Specifically, the  emission-line properties of these galaxies are well reproduced by models with metallicities $\zism\gtrsim0.004$, consistent with the results from previous studies \citep[e.g.,][]{brinchmann04, tremonti04,pacifici12}. Lower-metallicity galaxies are rare in the SDSS, but models with $0.0001\lesssim\zism\lesssim0.004$ will be crucial to interpret emission-line observations of chemically young galaxies at high redshifts, as already shown by \citet[][see also Section~\ref{sec:uv} below]{stark14,stark15a}. Fig.~\ref{fig:entire_grid} also confirms the usual finding that, observationally, nebular emission in metal-poor (metal-rich) galaxies seems to be associated with high (low) ionization parameter  \cite[e.g., CL01;][]{brinchmann04}.

It is important to stress that a same model can account simultaneously for the properties of a given galaxy in all panels of Fig.~\ref{fig:entire_grid}. We illustrate this by considering a `standard' model accounting roughly for the typical (i.e. median) properties of the observational sample in all panels (black cross in Fig.~\ref{fig:entire_grid}). This standard model has the following parameters: 
\begin{itemize}
\item $\zism=0.014$
\item $\xid=0.28$
\item $\log\Us=-3.4$
\item $\nh=10^2\,{\rm cm}^{-3}$
\item $\CO=\COsol=0.44$
\item $\mup=100\,\msol$.
\end{itemize}
We note that the metallicity $\zism=0.014$ is that closest to the present-day solar metallicity ($\zavsol=0.01524$) in Table~\ref{tab:parameters}. For reference, the standard model has a gas-phase oxygen abundance $12+\log\OHgas=8.70$, while the interstellar oxygen abundance is that corresponding to the metallicity $\zism=0.014$ in Table~\ref{tab:12logOH}, i.e., $12+\log\OH=8.80$.

\subsection{Influence of model parameters on optical emission-line properties}
\label{sec:optical_diagno}

We now briefly describe the influence of the main adjustable parameters of our model on the predicted optical emission-line properties of star-forming galaxies (see also CL01). In this description, we  explore the effect of varying a single parameter at a time, keeping the other main adjustable parameters fixed:

\begin{description}
\item {\textit{\textbf{Interstellar metallicity}}}. Fig.~\ref{fig:entire_grid} (solid line) shows that increasing \zism\ at fixed other parameters makes the \oiiihb\ ratio rise to a maximum (around $\zism\approx0.006$) and then decrease again. This is because gas cooling through collisionally excited optical transitions first increases as the abundance of metal coolants rises, until the electronic temperature drops low enough for cooling to become dominated by infrared fine-structure transitions \citep[e.g.,][]{spitzer78}. Efficient fine-structure cooling by doubly-ionized species in the inner parts of \hii\ regions also makes the \oiioiii\ ratio rise in Fig.~\ref{fig:entire_grid}d \citep{stasinska80}. The \siiha\ ratio behaves in a similar way to the \oiiihb\ ratio (Fig.~\ref{fig:entire_grid}b). In contrast, our inclusion of secondary nitrogen production causes the  \niiha\ and \niioii\ to rise steadily with metallicity (Figs~\ref{fig:entire_grid}a and \ref{fig:entire_grid}c). We note that, because of our adoption of the same metallicity for the stars and the ISM, lowering \zism\ also leads to a harder ionizing spectrum (since metal-poor stars evolve at higher effective temperatures than metal-rich ones; e.g., fig.~15 of \citealt{bressan12}). This has little influence on the results of Fig.~\ref{fig:entire_grid}, which are largely dominated by the other effects described above. 

\item {\textit{\textbf{Zero-age ionization parameter at the Str\"omgren radius}}}. Fig.~\ref{fig:entire_grid} shows that increasing \Us\ at fixed other parameters makes the \oiiihb\ ratio rise and the \oiioiii, \niiha\ and \siiha\ ratios drop. This is because increasing \Us\ at fixed density \nh\ and ionizing photon rate $Q(0)$ in our model amounts to increasing the effective gas filling factor $\epsilon$ (equation~\ref{eq:us} of Section~\ref{sec:photo_code}), causing the \hii\ regions to be more compact and concentrated close to the ionizing star clusters. This strengthens the high-ionization \oiii\ line relative to the lower-ionization \oiit, \nii\ and \siid\ lines.

\begin{figure}
\begin{center}
  \includegraphics[scale=0.43]{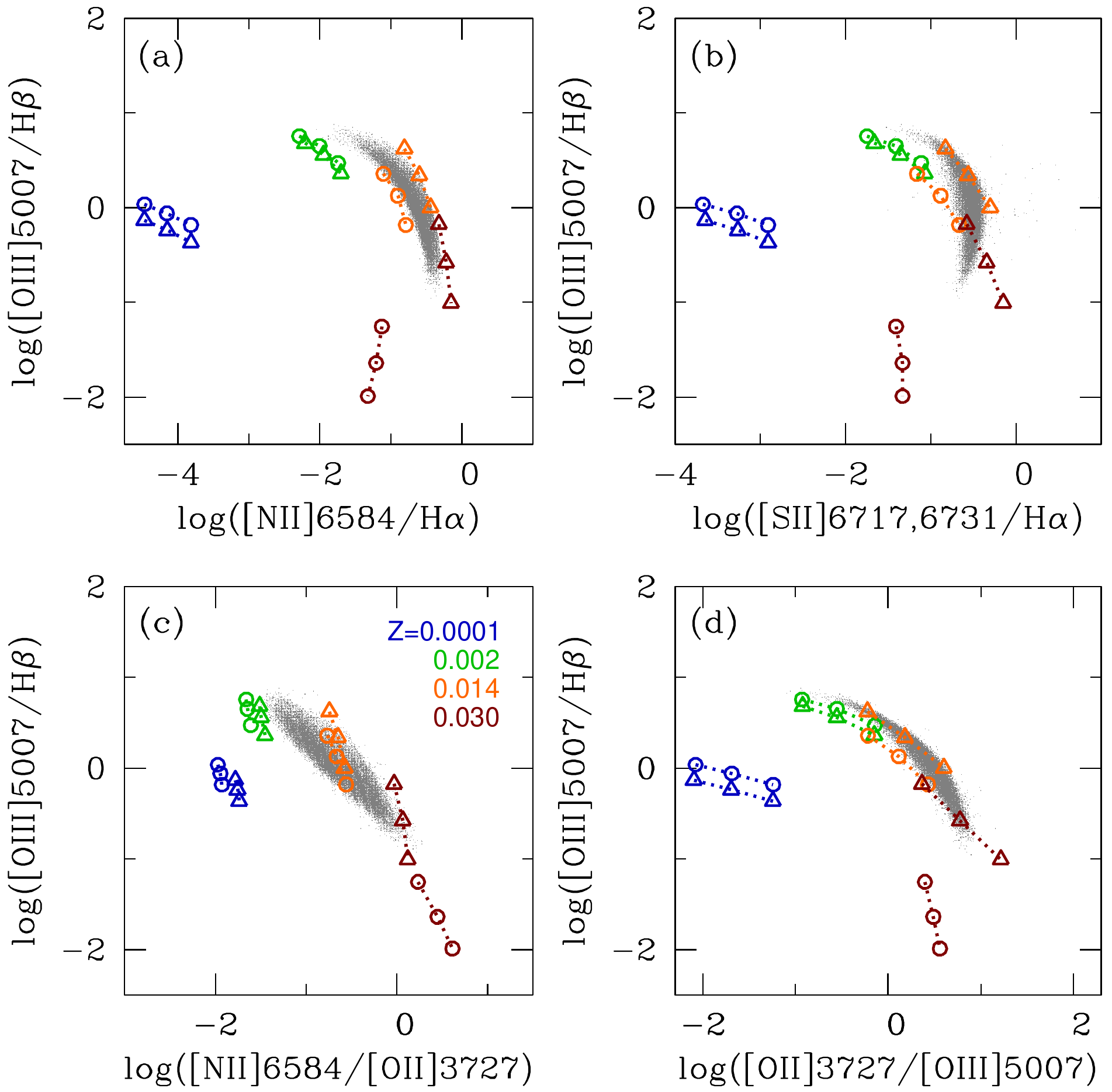}
  \caption[effect3]{Same as Fig.~\ref{fig:entire_grid}, but for a subset of interstellar metallicities (colour-coded as indicated) and associated zero-age ionization parameters (using the dependence of \Us\ on \zism\ identified by Carton et al., in preparation; see equation~4.2 of \citealt{chevallard16}): $\zism=0.0001$ for $\log\Us=-1.0$, $-1.5$ and $-2.0$ (in order of increasing \oiiihb\ ratio and connected by a line); $\zism=0.002$ for $\log\Us=-2.0$, $-2.5$ and $-3.0$; $\zism=0.014$ for $\log\Us=-2.5$, $-3.0$ and $-3.5$; and $\zism=0.030$ for $\log\Us=-3.0$, $-3.5$ and $-4.0$. In each panel, models are shown for two different dust-to-metal mass ratios, $\xid=0.1$ (circles) and 0.5 (triangles).}
  \label{fig:effect_xid_optic}
\end{center}
\end{figure}

\item {\textit{\textbf{Dust-to-metal mass ratio}}}. The effects of changes in \xid\ at fixed other parameters are shown in Fig.~\ref{fig:effect_xid_optic}. For clarity, we plot models for only a subset of 4 interstellar metallicities and 3 associated zero-age ionization parameters (using the dependence of \Us\ on \zism\ identified by Carton et al., in preparation; see equation~4.2 of \citealt{chevallard16}). Increasing \xid\ depletes metal coolants from the gas phase. The electronic temperature increases, as does cooling through collisionally excited optical transitions \citep[e.g.][]{shields95}. The implied rise in \niiha\ and \siiha\ ratios is significantly stronger than that in \oiiihb\ ratio (Figs~\ref{fig:effect_xid_optic}a and \ref{fig:effect_xid_optic}b), because oxygen is a refractory element strongly depleted from the gas-phase (making the \oiiihb\ ratio drop as \xid\ increases), while S and N are both non-refractory elements (Table~\ref{tab:abund_depl}). Fig.~\ref{fig:effect_xid_optic} further shows that, not surprisingly, the effect changing \xid\ is more pronounced at high than at low \zism. 

\begin{figure}
\begin{center}
  \includegraphics[scale=0.43]{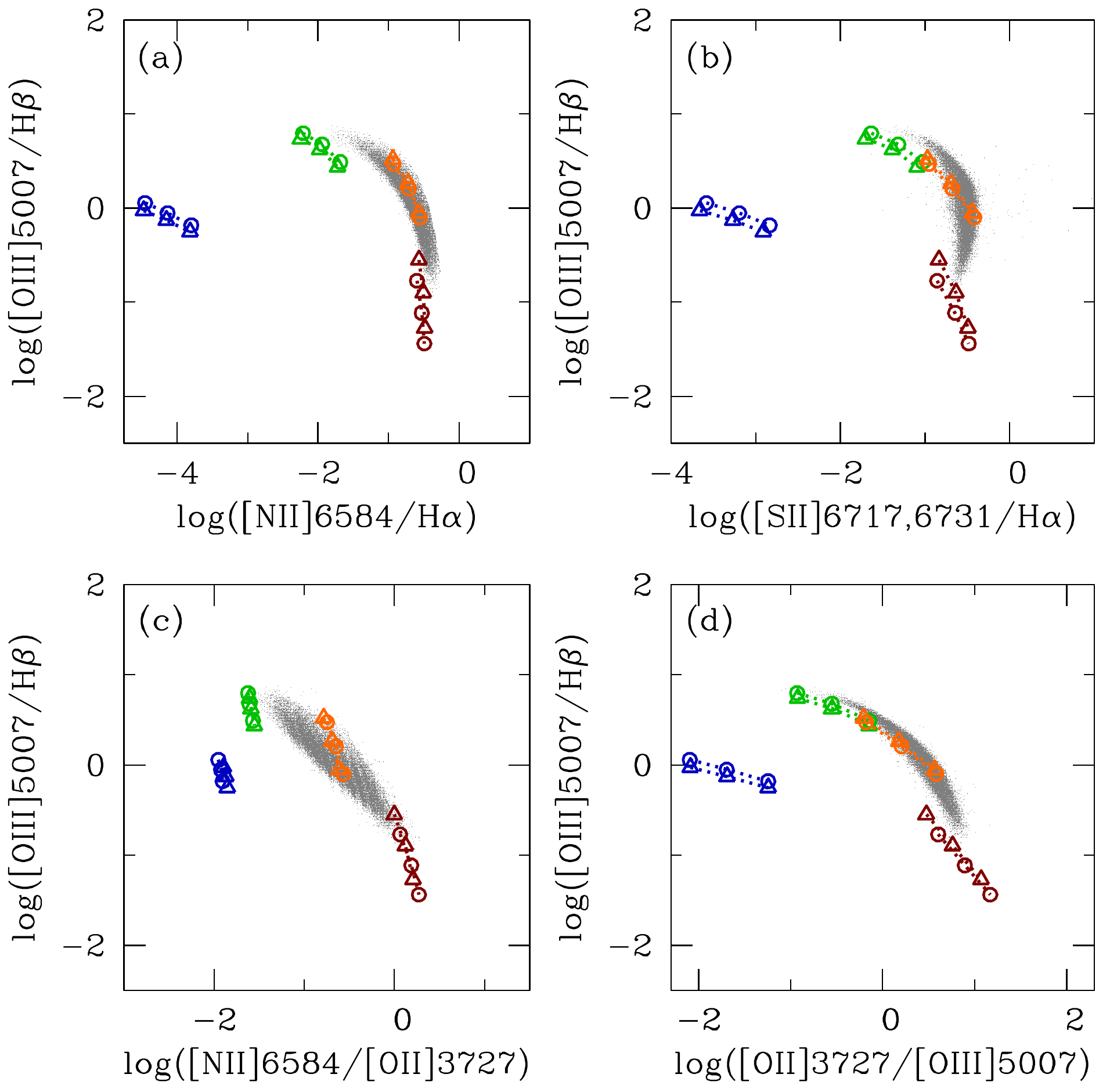}
\caption[effect3]{Same as Fig.~\ref{fig:effect_xid_optic}, but for models with two different carbon-to-oxygen ratios, $\CO=0.1$ (circles) and 1.4 (triangles) times \COsol.}
  \label{fig:effect_co_optic}
\end{center}
\end{figure}

\item {\textit{\textbf{Carbon-to-oxygen abundance ratio}}}. In our model, increasing the \CO\ ratio at fixed other parameters amounts to increasing the abundance of carbon and decreasing the abundances of all other heavy elements to maintain the same \zism\ (Section~\ref{sec:abund}). As Fig.~\ref{fig:effect_co_optic} shows, therefore, the effect of raising the \CO\ ratio on emission-line ratios involving N, O and S transitions is similar to that of lowering the interstellar metallicity \zism\ (shown by the solid line in Fig.~\ref{fig:entire_grid}).

\begin{figure}
\begin{center}
  \includegraphics[scale=0.43]{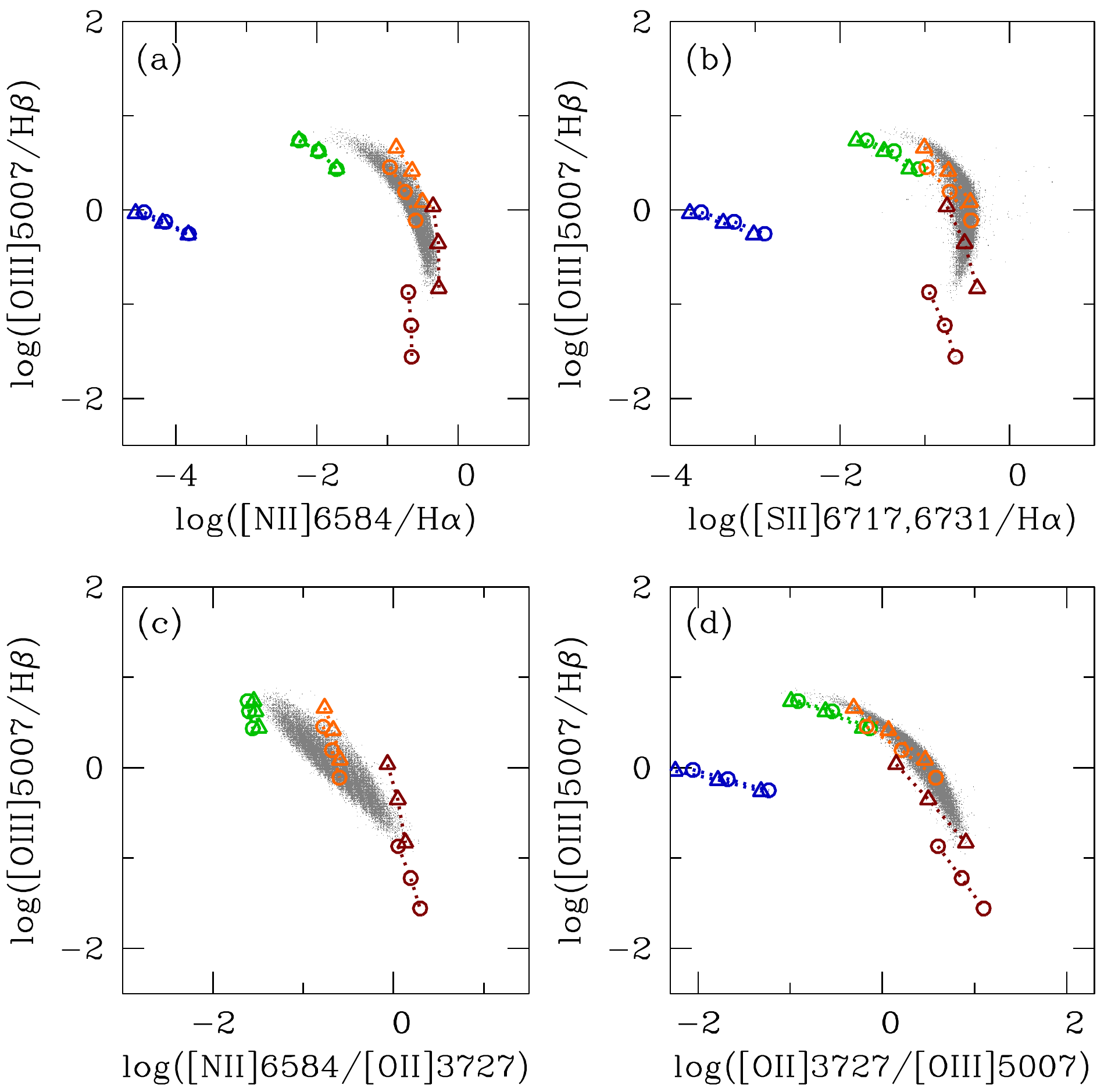}
  \caption[effect1]{Same as Fig.~\ref{fig:effect_xid_optic}, but for models with two different hydrogen densities,  $\nh=10\,{\rm cm}^{-3}$ (circles) and $10^3\,{\rm cm}^{-3}$ (triangles).}
  \label{fig:effect_nh_optic}
\end{center}
\end{figure}

\item {\textit{\textbf{Hydrogen gas density}}}. A rise in \nh\ increases the probability of an excited atom to be de-excited collisionally rather than radiatively. Since the critical density for collisional de-excitation is lower for infrared fine-structure transitions than for optical transitions, the net effect of raising \nh\ at fixed other parameters is to reduce the cooling efficiency through infrared transitions and increase that through optical transitions. Fig.~\ref{fig:effect_nh_optic} shows that the implied rise in the \oiiihb, \niiha\ and \siiha\ ratios is small at low metallicity, but much stronger at high metallicity, where infrared fine-structure transitions dominate the cooling \citep[e.g.,][see also the discussion of Fig.~\ref{fig:entire_grid} above]{oey93}.

\begin{figure}
\begin{center}
  \includegraphics[scale=0.43]{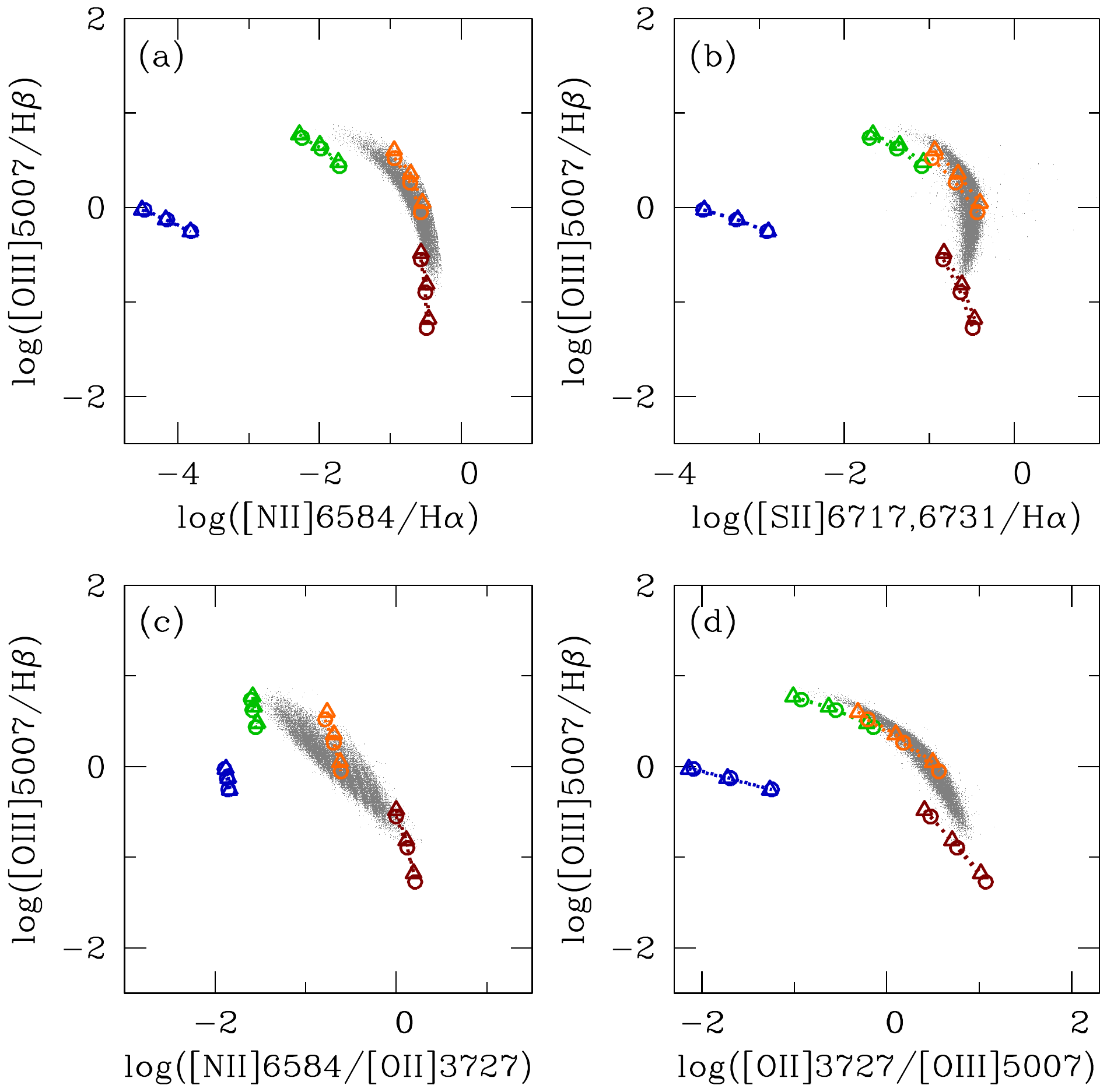}
  \caption[effect2]{Same as Fig.~\ref{fig:effect_xid_optic}, but for models with two different IMF upper mass cutoffs, $\mup=100\msol$ (circles) and $300\msol$ (triangles).}
  \label{fig:effect_mup_optic}
\end{center}
\end{figure}

\item {\textit{\textbf{Upper mass cutoff of the IMF}}}. Increasing \mup\ from 100 to 300\,\msol\ makes the ionizing spectrum of the stellar population harder, since stars with initial masses greater than 100\,\msol\ evolve at higher effective temperatures than lower-mass stars.  As Fig.~\ref{fig:effect_mup_optic} shows, this makes the high-ionization lines stronger, causing a rise of the \oiiihb\ ratio -- and to a lesser extent, the \niiha\ and \siiha\ ratios -- and a drop of the \oiioiii\ ratio.

\end{description}

Hence, the various adjustable parameters of the model of nebular emission described in Section~\ref{sec:modelling} influence, each in its own way, the optical emission-line spectra of star-forming galaxies. These specific signatures enable one to constrain simultaneously, in return, the star-formation and interstellar-gas parameters of observed galaxies with measured optical emission-line intensities \citep[see, e.g., CL01;][]{brinchmann04,pacifici12}. As mentioned previously, a main originality (other than the use of updated stellar population and photoionization prescriptions) of the models presented in Figs~\ref{fig:entire_grid}--\ref{fig:effect_mup_optic} above relative to previous models of the optical nebular emission from star-forming galaxies lies in the versatile, yet self-consistent, accounting of gas-phase versus interstellar element abundances, which allows investigations of chemically young galaxies with non-scaled solar element abundance ratios.

\section{Ultraviolet emission-line properties}
\label{sec:uv}

One of our primary motivations in this work is to build a library of photoionization models useful to interpret observations of the rest-frame ultraviolet emission from young star-forming galaxies at high redshifts. In this section, we investigate the ultraviolet properties of the grid of models presented in the previous sections and compare these predictions with available observations of a small, heterogeneous sample of local and distant star-forming galaxies. For illustration purposes, we have selected -- by means of a systematic investigation by eye of ratios involving the strongest ultraviolet emission lines -- a set of emission-line ratios most sensitive to changes in the adjustable parameters of our model. These ratios involve six emission lines (or multiplets) commonly detected in the spectra of star-forming galaxies: \nvd\ (hereafter \nv); \civd\ (hereafter \civt); \heii; \oiiis; \siliiid\ (hereafter \siliiit); and \ciiid\ (hereafter \ciiit). We note that, while the luminosities of \oiiis, \siliiit\ and \ciiit\ can usually be measured in a straightforward way when these lines are detectable, measurements of nebular \heii\ may be challenged by the presence of a broad component arising from Wolf-Rayet stars, especially at metallicities $\zism\gtrsim0.006$ (because of both a drop in photons capable of producing nebular \heii\ and a rise in stellar \heii\ emission as metallicity increases; e.g. \citealt{schaerer98}). At such metallicities, measurements of nebular \nv\ and \civt\ are even more challenging, because of the contamination by strong P-Cygni absorption features from O-star winds \citep{walborn84} coupled with interstellar absorption (see Vidal-Garc\'{\i}a et al., in preparation, for a detailed modelling of these competing effects).

\begin{figure*}
\begin{center}
  \includegraphics[scale=0.75]{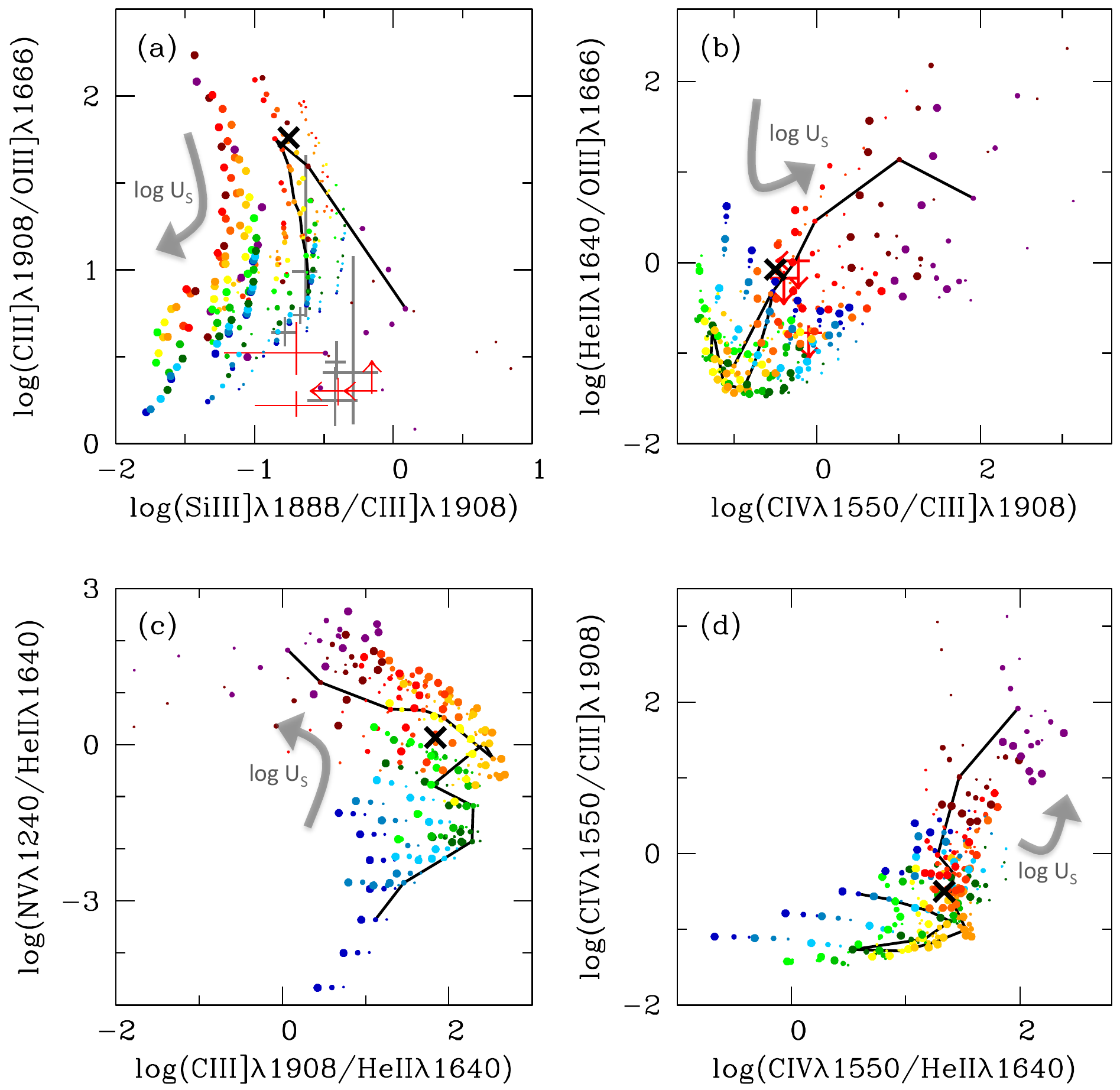}
  \caption[effect3]{Luminosity ratios of prominent ultraviolet emission lines predicted by the photoionization models described in Section~\ref{sec:modelling}: (a) \ciiit/\oiiis\ against \siliiit/\ciiit; (b) \heii/\oiiis\ against \civt/\ciiit; (c) \nv/\heii\ against \ciiit/\heii; and (d) \civt/\ciiit\ against \civt/\heii. In each panel, the models and solid line are the same as in Fig.~\ref{fig:entire_grid}. In panel (a), the grey and red crosses refer to observations (including error bars and upper limits) of, respectively, six giant extragalactic \hii\ regions in nearby low-luminosity, metal-poor, dwarf irregular galaxies observed with {\it HST}/FOS by \citet{garnett95} and four low-mass, gravitationally-lensed dwarf galaxies at redshift in the range $2\lesssim z\lesssim3$ observed with Keck/LRIS and VLT/FORS2 by \citet[][also reported in panel b]{stark14}.}
  \label{fig:entire_grid_uv}
\end{center}
\end{figure*}

Fig.~\ref{fig:entire_grid_uv} shows the ultraviolet properties of the same set of photoionization models (extracted from the grid of Table~\ref{tab:parameters}) as that for which we showed the optical properties in Fig.~\ref{fig:entire_grid}. Also shown in grey in the top-left panel of Fig.~\ref{fig:entire_grid_uv} are measurements of the \ciiioiii\ and \siliiiciii\ ratios in the spectra of six giant extragalactic \hii\ regions in nearby low-luminosity, metal-poor, dwarf irregular galaxies observed with the {\it Hubble Space Telescope}/Faint Object Spectrograph ({\it HST}/FOS; from \citealt{garnett95}). In the top two panels, the red data points show measurements of (and limits on) the \ciiioiii, \siliiiciii, \heiioiii\ and \civciii\ ratios in the spectra of four low-mass, gravitationally-lensed dwarf galaxies at redshift in the range $2\lesssim z\lesssim3$ observed with the Keck low-resolution imaging spectrometer (LRIS) and Very Large Telescope Focal Reducer and Low Dispersion Spectrograph (VLT FORS2; from \citealt{stark14}). The model grid encompasses the few observational measurements in Fig.~\ref{fig:entire_grid_uv}. In fact, the models presented here have already been used successfully to interpret rest-frame ultraviolet observations of the nebular emission from young star-forming galaxies at high redshifts \citep{stark14,stark15a,stark15b,stark16}.

We now explore the influence of the main adjustable parameters of our model on the predicted emission-line properties of star-forming galaxies in these four ultraviolet diagnostics diagrams. As in Section~\ref{sec:optical_diagno} above, we describe the effect of varying a single parameter at a time, keeping the other main adjustable parameters fixed:

\begin{description}
\item {\textit{\textbf{Interstellar metallicity}}}. Globally, the effect of increasing \zism\ at fixed other parameters, shown by the solid line in Fig.~\ref{fig:entire_grid_uv}, can be understood in terms of the balance pointed out in Section~\ref{sec:optical_diagno} (Fig.~\ref{fig:entire_grid}) between the implied rise in the abundance of coolants, associated drop in electronic temperature and cooling through infrared fine-structure transitions. As a result, ratios of metal-line to \heii\ luminosities tend to rise, stagnate and eventually drop again when metallicity increases in Fig.~\ref{fig:entire_grid_uv}, while the inclusion of secondary nitrogen production causes the \nvheii\ ratio to rise steadily with \zism\ (Fig.~\ref{fig:entire_grid_uv}c).

\item {\textit{\textbf{Zero-age ionization parameter at the Str\"omgren radius}}}. In our model, increasing $U_{\rm S}$ at fixed other adjustable parameters causes the \hii\ regions to be more compact and concentrated close to the ionizing star clusters (Section~\ref{sec:optical_diagno}). Overall, this makes the luminosity ratios of lines with highest ionization potential to lines with lower ionization potential (such as the inverse of the \ciiioiii\ ratio plotted in Fig.~\ref{fig:entire_grid_uv}a and the \civciii\ ratio) and the \heii\ line (such as the \civheii\ and \nvheii\ ratios) rise in Fig.~\ref{fig:entire_grid_uv} (see fig.~1 of \citealt{feltre16} for a graphical summary of the ionization potentials of the different species considered in Fig.~\ref{fig:entire_grid_uv}).

\begin{figure}
\begin{center}
  \includegraphics[scale=0.43]{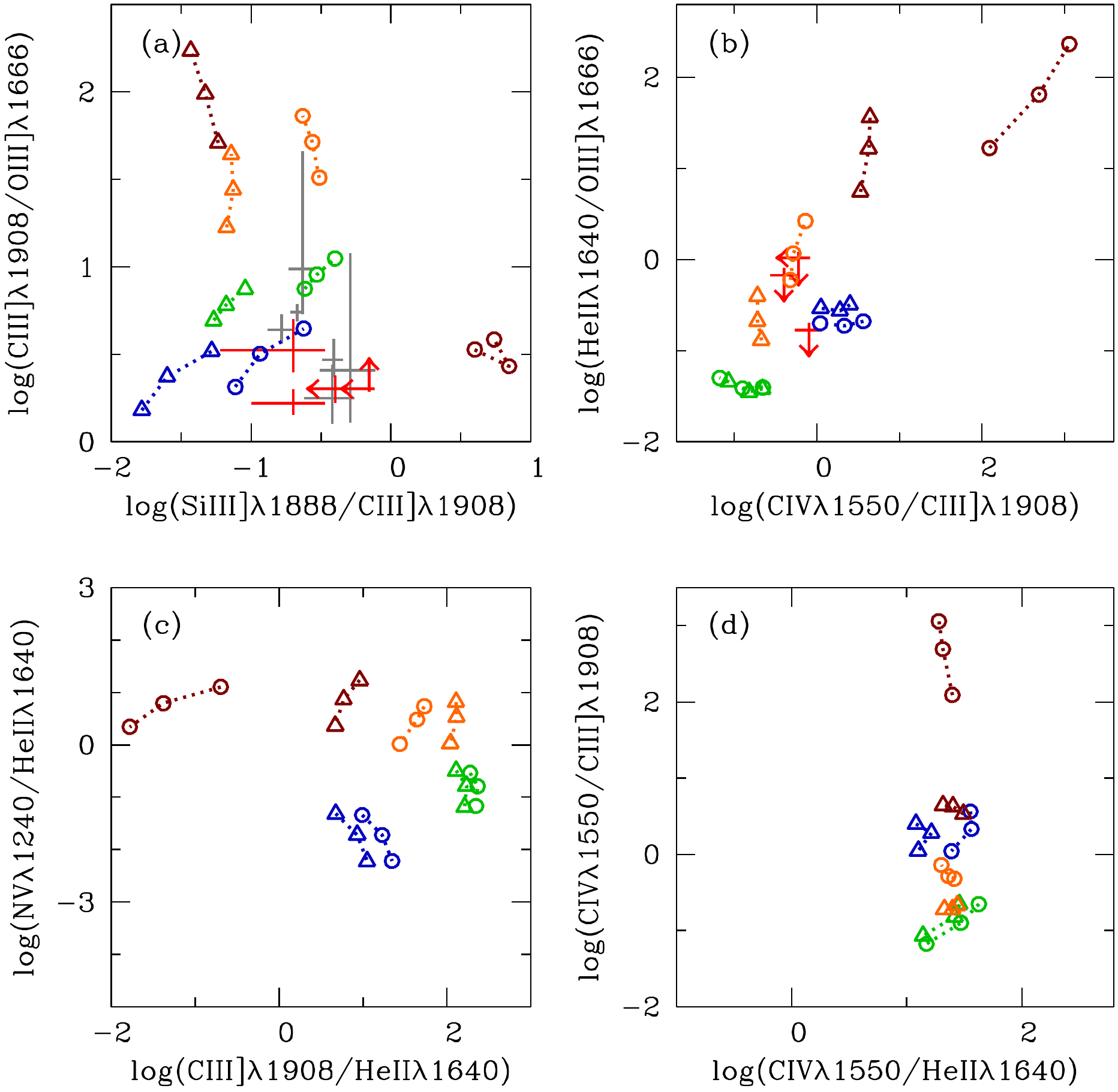}
  \caption[effect3]{Same as Fig.~\ref{fig:entire_grid_uv}, but for the same subset of interstellar metallicities and associated zero-age ionization parameters as in Fig.~\ref{fig:effect_xid_optic}. In each panel, models are shown for two different dust-to-metal mass ratios, $\xid=0.1$ (circles) and 0.5 (triangles).}
  \label{fig:effect_xid_uv}
\end{center}
\end{figure}

\item {\textit{\textbf{Dust-to-metal mass ratio}}}. Fig.~\ref{fig:effect_xid_uv} illustrates the influence of \xid\ on the predicted ultraviolet emission-line ratios, showing for clarity only models for the same subset of \zism\ and \Us\ combinations as used in Fig.~\ref{fig:effect_xid_optic} above. As for optical transitions (Section~\ref{sec:optical_diagno}), the response of ultraviolet transitions to a rise in \xid\ results from a balance between the implied depletion of coolants from the gas phase, the associated rise in electronic temperature and the relative depletions of different species (Table~\ref{tab:abund_depl}). For example, the \ciiiheii\ and \civheii\ ratios drop at low \zism\ as \xid\ rises, because of the disappearance of C from the gas phase, but the trend is opposite at high \zism, because the rise in electronic temperature induced by the depletion of heavy elements is more significant (Figs~\ref{fig:effect_xid_uv}c and \ref{fig:effect_xid_uv}d). In Fig.~\ref{fig:effect_xid_uv}a, the \siliiiciii\ drops at all metallicities when \xid\ rises, because Si is far more depleted than C from the gas phase. In contrast, since N is not depleted, the \nvheii\ ratio shows only a mild increase as \xid\ rises in Figs~\ref{fig:effect_xid_uv}c, because of the rise in electronic temperature. We note that, at high metallicity especially, the increase in dust optical depth ($\tau_{\rm d}\propto \xid \zism\nh\epsilon$) induced by a rise in \xid\ makes the electronic temperature drop through the enhanced absorption of energetic photons, causing the \civciii\  and \civheii\ ratios to drop and the \ciiioiii\ ratio to rise.  

\begin{figure}
\begin{center}
  \includegraphics[scale=0.43]{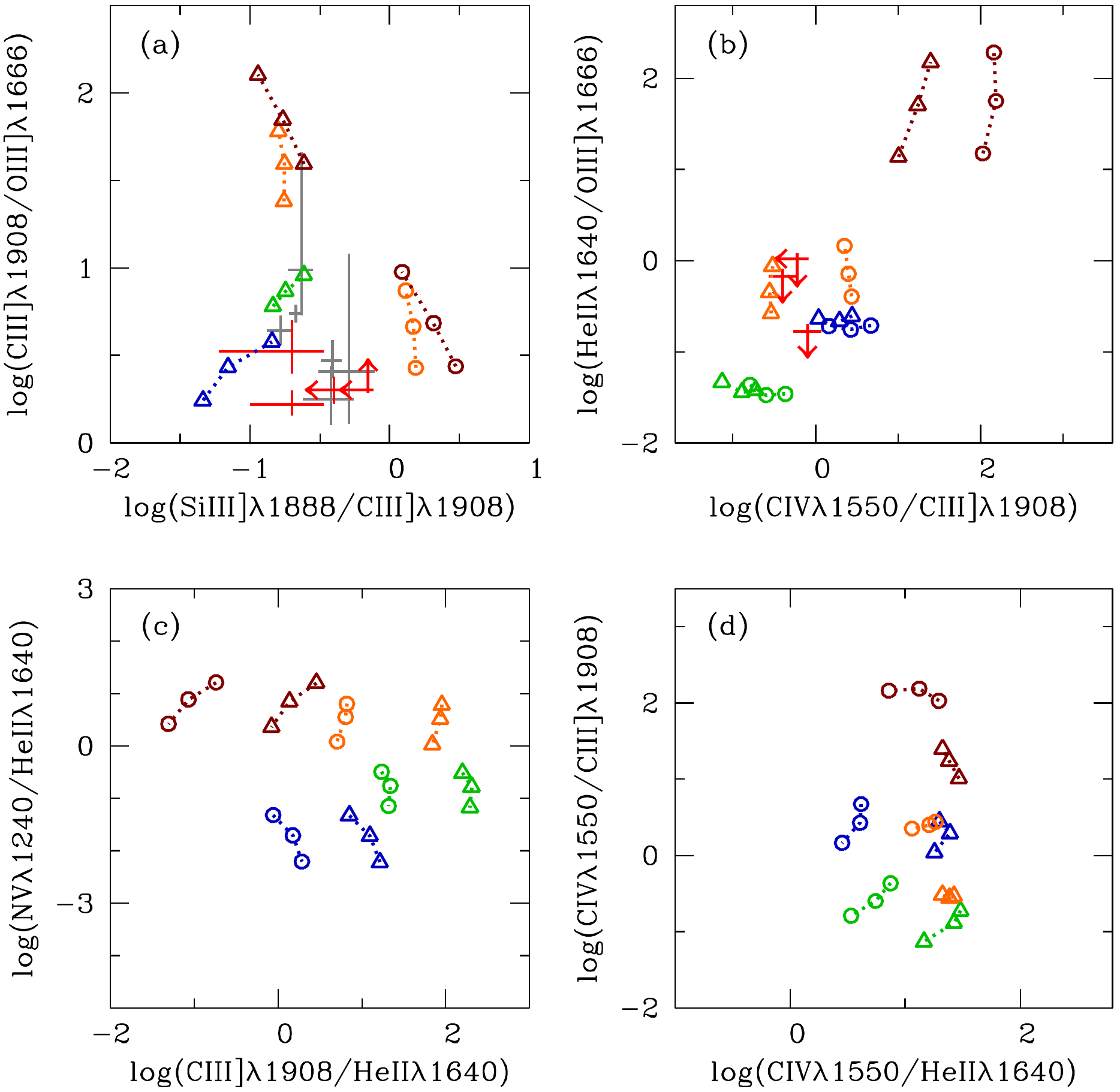}
  \caption[effect2]{Same as Fig.~\ref{fig:effect_xid_uv}, but for models with two different carbon-to-oxygen ratios, $\CO=0.1$ (circles) and 1.4 (triangles) times \COsol.}
  \label{fig:effect_co_uv}
\end{center}
\end{figure}

\item {\textit{\textbf{Carbon-to-oxygen abundance ratio}}}.  Increasing the \CO\ ratio at fixed other parameters consists in increasing the abundance of C while decreasing those of all other heavy elements (Section~\ref{sec:optical_diagno}). Fig.~\ref{fig:effect_co_uv} shows that, as a result, raising the \CO\ ratio makes the \ciiioiii, \ciiiheii\ and \civheii\ ratios markedly larger and the \siliiiciii\ ratio markedly smaller. Another important conclusion we can draw from this figure is that the presence of the \ciiit, \civt\ and \oiiit\ emission lines at ultraviolet wavelengths makes ultraviolet-line ratios more direct tracers of the \CO\ ratio of young star-forming galaxies than the standard optical emission lines investigated in Fig.~\ref{fig:effect_co_optic}.

\begin{figure}
\begin{center}
  \includegraphics[scale=0.43]{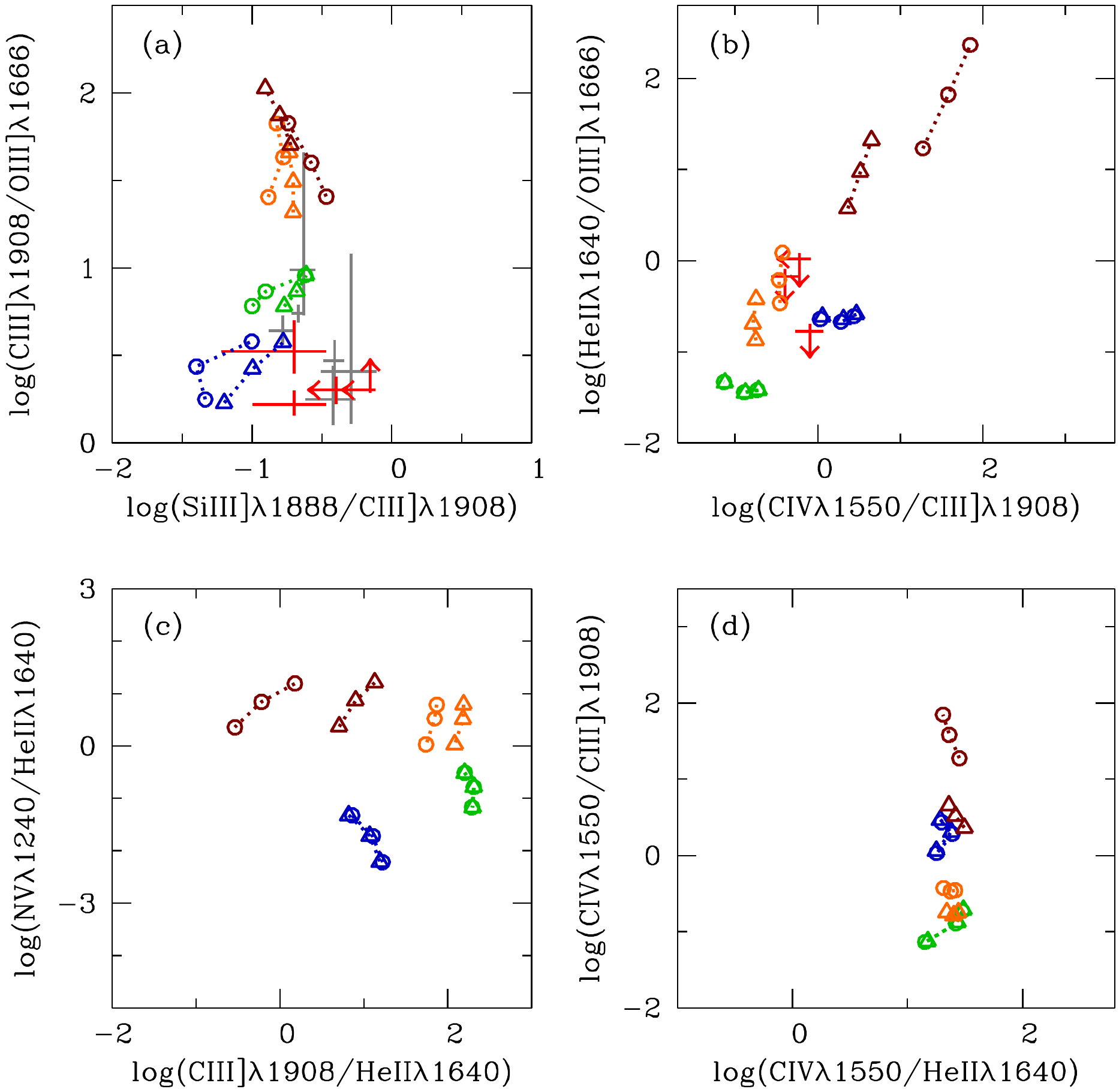}
  \caption[effect1]{Same as Fig.~\ref{fig:effect_xid_uv}, but for models with two different hydrogen densities,  $\nh=10\,{\rm cm}^{-3}$ (circles) and $10^3\,{\rm cm}^{-3}$ (triangles).}
  \label{fig:effect_nh_uv}
\end{center}
\end{figure}

\item {\textit{\textbf{Hydrogen gas density}}}.  The effect of raising \nh\ at fixed other parameters is to increase radiative cooling through ultraviolet and optical transitions relative to infrared ones (Section~\ref{sec:optical_diagno}). As Fig.~\ref{fig:effect_nh_uv} shows, this makes the \ciiiheii\ and \civheii\ ratios rise and the \heiioiii\ ratio drop. Also, in our model, increasing $n_{\rm H}$ at fixed other parameters implies lowering the volume-filling factor as $\epsilon\propto1/\sqrt{n_{\rm H}}$ (equation \ref{eq:us}). As a result, the dust optical depth ($\tau_{\rm d}\propto \xid \zism\nh\epsilon$) rises as $\sqrt{\nh}$, increasing the absorption of energetic photons \citep[see also][]{feltre16}, and hence, the \civciii\ ratio drops (Figs~\ref{fig:effect_nh_uv}b and \ref{fig:effect_nh_uv}d). Similarly to what was noted about optical transitions (Fig.~\ref{fig:effect_nh_optic}), the influence of \nh\ on ultraviolet emission-line ratios in Fig.~\ref{fig:effect_nh_uv} is weaker at low than at high metallicity,  where infrared fine-structure transitions usually dominate the cooling.

\begin{figure}
\begin{center}
  \includegraphics[scale=0.43]{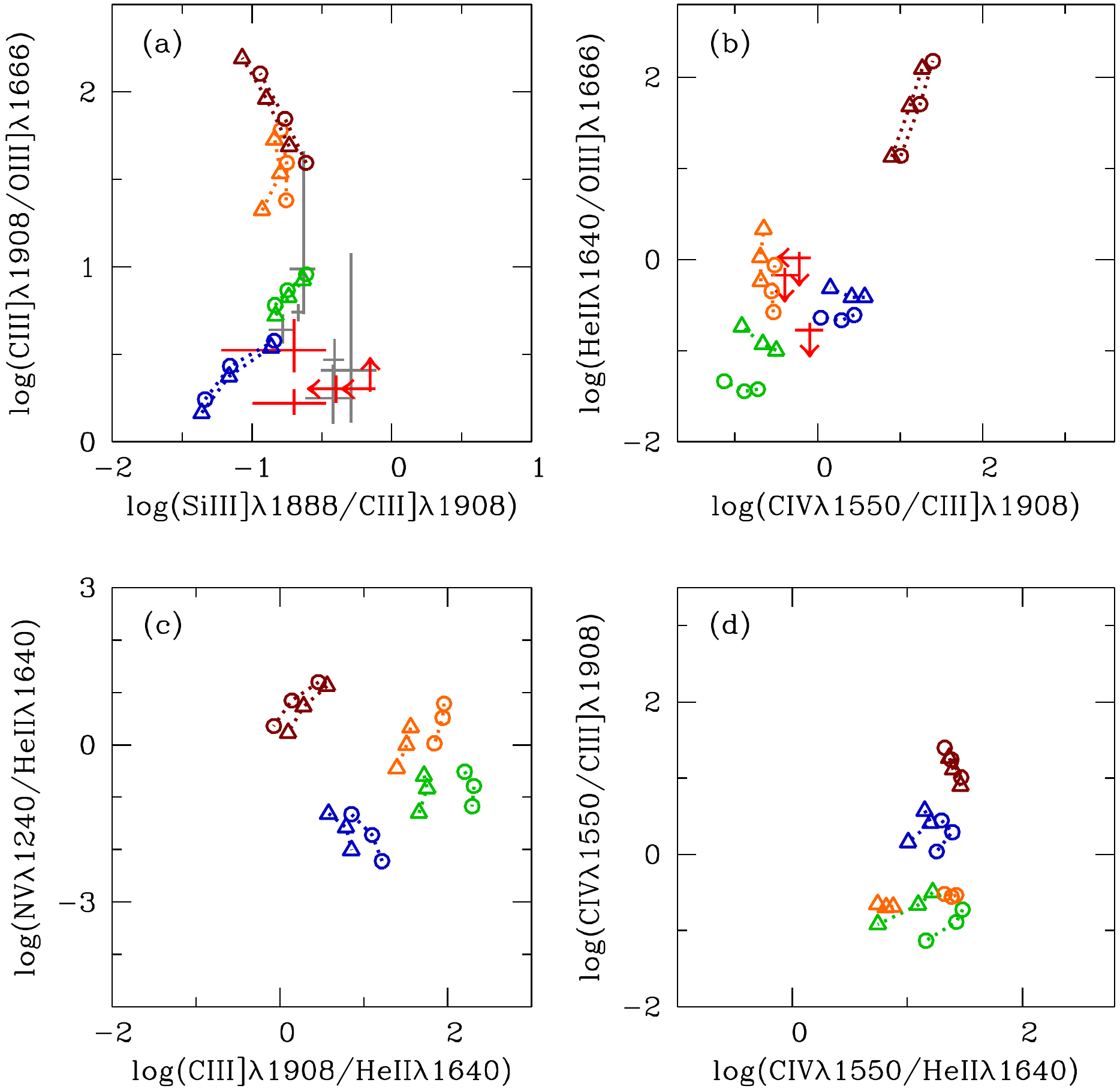}
  \caption[effect2]{Same as Fig.~\ref{fig:effect_xid_uv}, but for models with two different IMF upper mass cutoffs, $\mup=100\msol$ (circles) and $300\msol$ (triangles).}
  \label{fig:effect_mup_uv}
\end{center}
\end{figure}

\item {\textit{\textbf{Upper mass cutoff of the IMF}}}. A rise in \mup\ from 100 to 300\,\msol\ makes the ionizing spectrum harder (Section~\ref{sec:optical_diagno}), the signature of which is much stronger at ultraviolet than at optical wavelengths. Fig.~\ref{fig:effect_mup_uv} (to be compared with Fig.~\ref{fig:effect_mup_optic}) shows that a dominant effect at far-ultraviolet wavelengths is to boost the luminosity of the \heii\ recombination line, making the \heiioiii\ ratio rise and the \ciiiheii, \civheii\ and \nvheii\ ratios drop significantly. Collisionally excited metal transitions are also affected, in the sense that high-ionization lines become more prominent. This is the reason for the slight drop in \siliiiciii\ and \ciiioiii\ ratios and rise in the \civciii\ ratio.
\end{description}

In summary, by comparing Figs~\ref{fig:entire_grid}--\ref{fig:effect_mup_optic} with Figs~\ref{fig:entire_grid_uv}--\ref{fig:effect_mup_uv}, we may conclude from our investigation in this section and the previous one that ultraviolet emission-line ratios are significantly more sensitive than optical ones to some parameters of young star-forming galaxies, such as the carbon-to-oxygen abundance ratio, hydrogen gas density, upper mass cutoff of the IMF and even dust-to-metal mass ratio. This makes ultraviolet emission lines ideal tracers of the early chemical evolution of galaxies, provided that these lines, which are typically much fainter than their optical counterparts, can be detected. We stress that our ability to draw such conclusions relies largely on the incorporation in our model of a versatile and self-consistent treatment of gas-phase versus interstellar element abundances.

\section{Limitations of standard methods of abundance measurements}
\label{sec:icf}
\vspace{3mm} 

In Sections~\ref{sec:optical_diagno} and \ref{sec:uv} above, we have described the dependence of the ultraviolet and optical nebular emission of a galaxy on physical properties. The specific dependence of each feature on different physical parameters enables one to extract valuable constraints on star-formation and chemical-enrichment properties from observed emission-line fluxes. This can be achieved by combining the photoionization models of star-forming galaxies described in Section~\ref{sec:modelling} with a sophisticated spectral interpretation tool, as proposed for example in the framework of the \beagle\ tool by \citet[][see their section 2.3]{chevallard16}. As a complement to such studies, in the present section, we explore how the photoionization model of Section~\ref{sec:modelling} can also help identify potential limitations of abundance measurements involving the more approximate, although widely used, `direct-\Te' method.

\subsection{The `direct-\Te' method}
\label{sec:directTe}

In the absence of sophisticated modeling, the best measurements of chemical abundances from nebular emission lines are generally considered to be those combining an estimate of the electron temperature (\Te) from ratios of auroral to nebular forbidden-line intensities (such as \oiiir\ or \niir) with an estimate of the electron density ($\nnee$$\sim$$\nh$ in the ionized gas) from ratios of nebular forbidden-line intensities (such as \oiir\ or \siir; see, e.g., \citealt{aller59,peimbert67,peimbert69,aller84}).\footnote{We use the standard nomenclature and refer to forbidden transitions from the first-excited to ground levels as `nebular lines', and to those from the second- to first-excited levels as `auroral lines'.}  We note that a usually neglected uncertainty in this approach is that the ions used to derive \Te\ and \nnee\ often trace different ionized zones \citep[e.g., O$^{2+}$ versus O$^{+}$; see][]{keenan00}. Once \Te\ is estimated for some ionic species (e.g. O$^{2+}$), the total gas-phase abundance of the corresponding element (in this case, O) can be inferred by adding the contributions by the other stages of ionization (here mainly O$^{+}$ and O$^{0}$). This is generally achieved by using the observed luminosities of corresponding emission lines (e.g., \oiit, \oi) and by appealing to photoionization models to estimate the associated electronic temperatures, based on the one measured in the first ionization zone \citep[e.g., equation 6 of][]{pagel92,kobulnicky96}. Photoionization models must also be invoked to compute the gas-phase abundances of other heavy elements (e.g., C, N, S), which requires estimates of the electron temperatures pertaining to the zones populated by the corresponding ions \citep[e.g.,][]{garnett92,izotov99,stasinska05,izotov06}.

Hence, in the end, the `direct-\Te' method of abundance measurements relies indirectly on photoionization calculations. This implies that element abundances derived using this method are tied to the  physical parameters of the specific models adopted (abundances, depletion, ionizing radiation, gas density, etc.). As a result, standard calibrations of the direct-\Te\ method, based on observations of nearby \hii\ regions and galaxies, may not be fully appropriate to investigate the abundances of, for example, chemically young galaxies in the early Universe (Section~\ref{sec:intro}). Moreover, inferences made using the direct-\Te\ method generally do not provide any insight into interstellar (as opposed to gas-phase) element abundances in the ionized gas (Section~\ref{sec:abunddepl}). The models presented in Sections~\ref{sec:modelling}--\ref{sec:uv} above offer a unique means of investigating the impact of these limitations on abundance measurements in chemically young star-forming galaxies. In the remainder of this section, we focus on measurements of the \CO\ ratio, which appears to be a particularly sensitive probe of cosmic chemical evolution (e.g., \citealt{erb10,cooke11}; see also \citealt{garnett99}).

\subsection{A case study: the \CO\ ratio}
\label{sec:COicf}

\begin{figure*}
\begin{center}
   \includegraphics[scale=0.75]{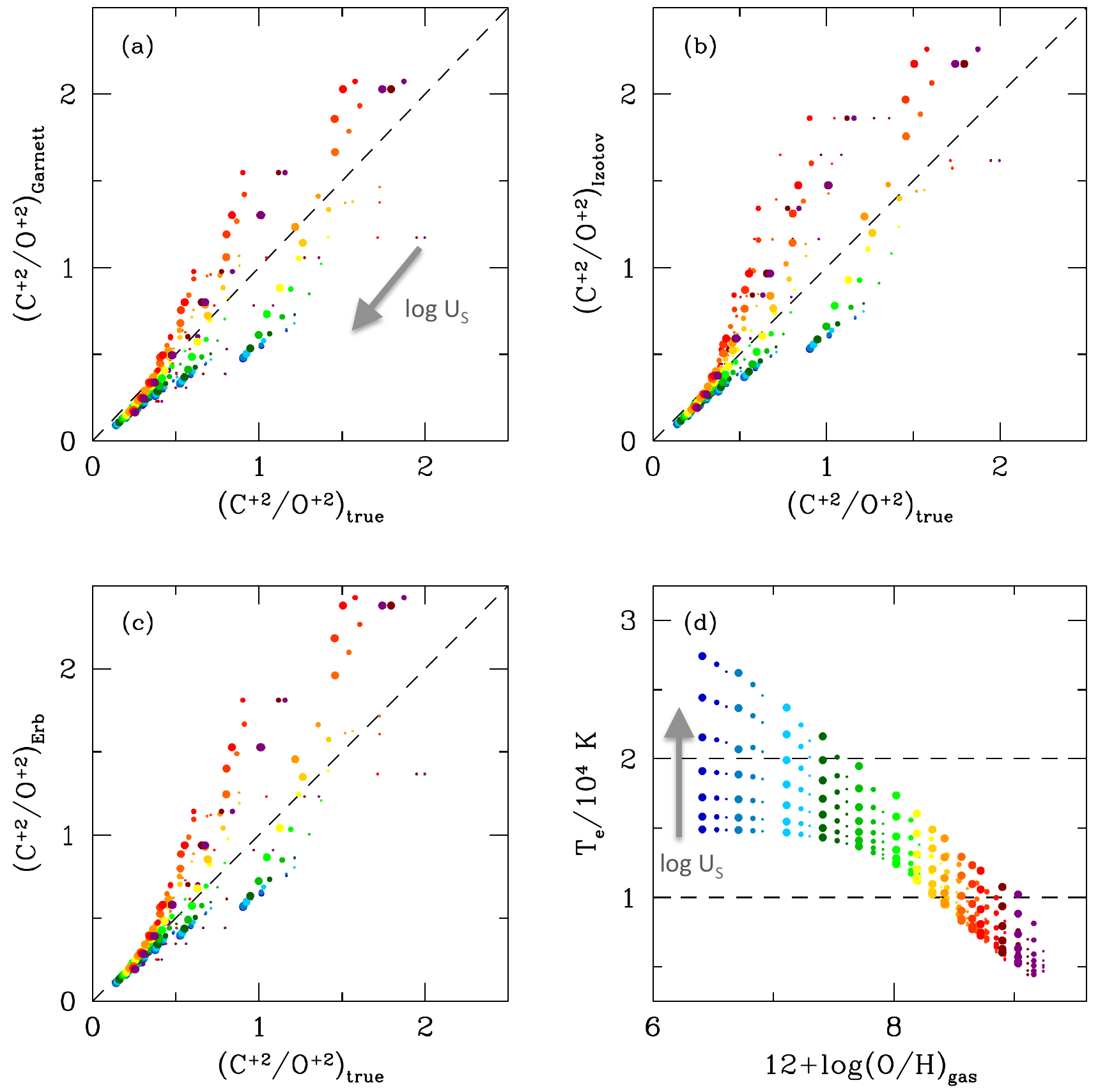}
  \caption[effect3]{\CppOpp\ ionic abundance ratio estimated from emission-line luminosities via standard formulae involving the direct-\Te\ method plotted against true \CppOpp\ ratio, for the same models as in Fig.~\ref{fig:entire_grid}: (a) according the prescription of \citet[][equation~\ref{eq:eq_c3o3_garnett} of Section~\ref{sec:COicf}]{garnett95icf}; (b) according the prescription of \citet[][equation~\ref{eq:eq_c3o3_izotov}]{izotov99}; (c) according the prescription of \citet[][equation~\ref{eq:co_from_icfr}]{erb10}. (d) Model electronic temperature in the O$^+2$ plotted against gas-phase oxygen abundance for the same models as in panels~(a)--(c). Dashed horizontal lines brackett the approximate \Te\ range over which equations~\eqref{eq:eq_c3o3_garnett}--\eqref{eq:eq_c3o3_erb} were calibrated (see text for details).}
  \label{fig:c3o3_all}
\end{center}
\end{figure*}

The \CO\ ratio of star-forming galaxies is often estimated through measurements of C$^{2+}$ and O$^{2+}$ emission lines, using the direct-\Te\ method and an ionization correction factor, as outlined in Section~\ref{sec:directTe} above \citep[e.g.,][]{garnett92,garnett95icf,izotov99,erb10}. This is traditionally expressed as
\begin{equation}
{\rm \left(\frac{C}{O}\right)_\mathrm{gas}= \frac{C^{+2}}{O^{+2}}} \times \left[\frac{X(\rm{C}^{+2})}{X(\rm{O}^{+2})}\right]^{-1}= {\rm\frac{C^{+2}}{O^{+2}} \times ICF,}
\label{eq:co_from_icfr}
\end{equation}
where $X(\rm{C}^{+2})\equiv{\rm C^{+2}/C_\mathrm{gas}}$ and $X(\rm{O}^{+2})\equiv{\rm O^{+2}/O_\mathrm{gas}}$ are the volume-averaged fractions of doubly-ionized C and O in the gas phase. Here, we have made explicit use of a subscript `gas' to emphasize the fact that equation~\eqref{eq:co_from_icfr} is generally used to investigate the gas-phase \CO\ ratio. We now use our model of nebular emission from star-forming galaxies to investigate the uncertainties affecting both factors in the right-hand side of this expression: the conversion of emission-line luminosities into a C$^{2+}$/O$^{2+}$ ratio; and the dependence of the ICF on photoionization conditions. We also explore below the relation between the \CppOpp\ ratio and the  interstellar (i.e. gas+dust-phase) \CO\ ratio in our model.

Several formulae have been proposed to convert measurements of ultraviolet and optical C$^{2+}$ and O$^{2+}$ emission-line luminosities into a ratio of (volume-averaged) densities of C$^{2+}$ to O$^{2+}$, based on theoretical computations of the associated collision strengths and emission-rate coefficients, combined with the direct-\Te\ method \citep[e.g., chapter 5 of][]{aller84}. These include the prescription of \citet{garnett95icf},\footnote{The flux labelled $I(\lambda1909)$ in equation~(2) of \citet{garnett95icf} refers to the \ciiit\ doublet, not resolved in the {\it HST}/FOS observations.}
\begin{equation}
\left(\frac{\rm C^{+2}}{\rm O^{+2}}\right)_{\rm Garnett} = 0.089\; e^{-1.09/\te}\, \frac{L(\ciiit)}{L(\oiiis)}\,,
\label{eq:eq_c3o3_garnett}
\end{equation}
that of \citet[][see also \citealt{aller84}]{izotov99},
\begin{equation}
\left(\frac{\rm C^{+2}}{\rm O^{+2}}\right)_{\rm Izotov} = 0.093\; e^{4.656/\te}\, \frac{L(\ciiit)}{L(\oiiitopt)}\,,
\label{eq:eq_c3o3_izotov}
\end{equation}
and that of \citet[][see also \citealt{shapley03}]{erb10},
\begin{equation}
\left(\frac{\rm C^{+2}}{\rm O^{+2}}\right)_{\rm Erb} = 0.15\; e^{-1.1054/\te}\, \frac{L(\ciiit)}{L(\oiiid)}\,,
\label{eq:eq_c3o3_erb}
\end{equation}
where $\te = \Te/ 10^4\,$K and $L$ is the line luminosity. The electronic temperature in these expressions is usually taken to be that of the O$^{+2}$ zone. This can be estimated from the \oiiitopt/\oiiiaur\ luminosity ratio using the calibration of \citet{aller84},
\begin{equation}
\te=\frac{1.432}{\log[{L(\oiiitopt)}/{L(\oiiiaur)}]-\log C_T},
\label{eq:eq_te_aller}
\end{equation}
where 
\begin{equation}
C_T=\left(8.44-1.09\,\te+0.5\,\te^2-0.08\,\te^3\right)\,\frac{1+0.0004\,x}{1+0.044\,x}
\label{eq:eq_ct_aller}
\end{equation}
and $x = 10^{-4} \, (\nnee/{\rm cm}^{-3})\te^{-1/2}$. If the \oiiiaur\ line is not available (as in \citealt{erb10}), \te\ can be estimated instead from the \oiiid/\oiii\ luminosity ratio, using the calibration of \citet[][their fig.~1]{villar04}.

In Figs~\ref{fig:c3o3_all}a--\ref{fig:c3o3_all}c, we compare the \CppOpp\ ratios estimated using equations~\eqref{eq:eq_c3o3_garnett}--\eqref{eq:eq_c3o3_erb} from the emission-line properties of the same subset of photoionization models as in Figs~\ref{fig:entire_grid} and \ref{fig:entire_grid_uv} above (extracted from the grid of Table~\ref{tab:parameters}) with the true  \CppOpp\ ratios of these models (as provided by the \cloudy\ code). In practice, we solve for \te\ in equation~\eqref{eq:eq_te_aller} by setting \nnee\ equal to the hydrogen density $\nh=100\,{\rm cm}^{-3}$ of these models when evaluating $x$ in equation~\eqref{eq:eq_ct_aller}. The resulting electronic temperatures are shown as a function of gas-phase oxygen abundance, $12+\log\OHgas$, in Fig.~\ref{fig:c3o3_all}d. As expected, at fixed other model parameters, \Te\ drops markedly as the abundance of coolants rises (Sections~\ref{sec:optical_diagno} and \ref{sec:uv}).

Figs~\ref{fig:c3o3_all}a--\ref{fig:c3o3_all}c show that the \CppOpp\ ratio estimated using the standard formulae of equations~\eqref{eq:eq_c3o3_garnett}--\eqref{eq:eq_c3o3_erb} can differ substantially from the true one, by an amount ranging from a factor of $\sim$2 underestimate at low metallicity (blue points) to a factor of $\sim$2 overestimate at high metallicity (red points). The difference shrinks around solar metallicity (orange points). The most likely reason for the discrepancy at non-solar metallicities is that the formulae in equations~\eqref{eq:eq_c3o3_garnett}--\eqref{eq:eq_c3o3_erb} were derived using C$^{+2}$ ($^1S$--$^3P$) and O$^{+2}$ ($^3P$--$^5S_2$) collision strengths computed for a restricted range of electronic temperatures typical of \hii\ regions with solar-to-slightly sub-solar metallicity, i.e. $10,000\lesssim\Te/{\rm K}\lesssim20,000$ \citep{aller84,garnett95}. As Fig.~\ref{fig:c3o3_all}d shows, \hii\ regions at low and high metallicities can reach temperatures outside this range. Hence, we conclude from Fig.~\ref{fig:c3o3_all} that the ability to estimate the \CppOpp\ ratio from emission-line luminosities using simple standard recipes breaks down at non-solar metallicities.

\begin{figure}
\begin{center}
  \includegraphics[scale=0.65]{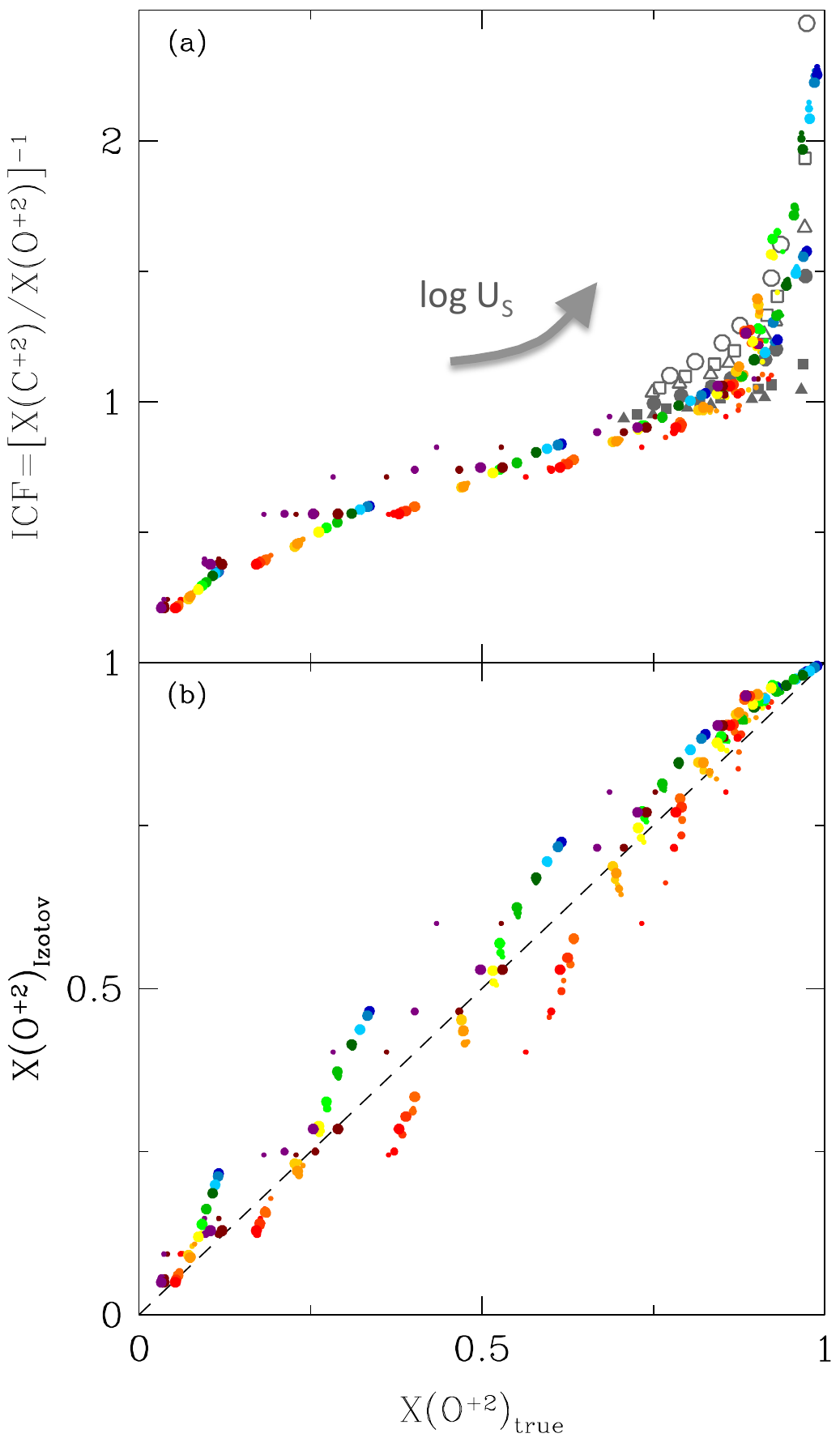}
  \caption{(a) Ionization correction factor entering equation~\eqref{eq:co_from_icfr}, $[X(\rm{C}^{+2})/X(\rm{O}^{+2})]^{-1}$, plotted against volume-averaged fraction of doubly-ionized oxygen, $X(\rm{O}^{+2})$, for the same models as in Fig.~\ref{fig:entire_grid}. The circles, squares and triangles show the results of early calculations by \citet{garnett95} for OB stellar associations of metallicity 0.001, 0.004 and 0.008, respectively, and ages 0\,Myr (filled symbols) and 2\,Myr (open symbols; see text for details). (b) $X(\rm{O}^{+2})$ estimated from emission-line luminosities via standard formulae involving the direct-\Te\ method (equations~\ref{xo}--\ref{oiii}) plotted against true $X(\rm{O}^{+2})$, for the same models as in (a).}
  \label{fig:icf_obs_param}
\end{center}
\end{figure}

We now turn to the other main potential source of uncertainty affecting estimates of the \COgas\ ratio via emission-line luminosities, i.e., the correction factor (ICF) to convert \CppOpp\ into \COgas\ in equation~\eqref{eq:co_from_icfr}. Fig.~\ref{fig:icf_obs_param}a shows the ICF as a function of volume-averaged fraction of doubly-ionized oxygen, $X(\rm{O}^{+2})$, for the same models of star-forming galaxies as in Fig.~\ref{fig:c3o3_all}. The quantity $X(\rm{O}^{+2})$ and the ICF depend sensitively on the ionization parameter and, at fixed \Us, the metallicity \zism, which influences both the electronic temperature and the hardness of the ionizing radiation (since $\zav=\zism$; see Sections~\ref{sec:optical_diagno} and \ref{sec:uv}). Specifically, a rise in \Us\ and a drop in \zism\ both make the fraction of doubly-ionized oxygen larger. At the same time, the lower ionization potential of C$^{+2}$ (47.9\,eV) relative to O$^{+2}$ (54.9\,eV) implies that $X(\rm{C}^{+2})$ starts to drop before $X(\rm{O}^{+2})$, because of the ionization of C to C$^{+3}$, causing the ICF to also increase. Also shown for comparison in Fig.~\ref{fig:icf_obs_param}a are the pioneer, dust-free calculations of the ICF by \citet{garnett95} for OB stellar associations at ages $t^\prime=0$ and 2\,Myr (in the notation of Section~\ref{sec:photo_code}), based on early prescriptions by \citet{mihalas72} and \citet{panagia73}. These calculations for large $X(\rm{O}^{+2})$ are in reasonable agreement with our more sophisticated models of \hii-region populations in star-forming galaxies.

It is important to note that, although the scatter in the ICF predicted at fixed $X(\rm{O}^{+2})$ by our models is moderate in Fig.~\ref{fig:icf_obs_param}a, the influence of this scatter on estimates of the \COgas\ ratio via equation~\eqref{eq:co_from_icfr} is amplified by the uncertainties affecting observational estimates of $X(\rm{O}^{+2})$ from ratios of oxygen emission lines through the direct-\Te\ method (Section~\ref{sec:directTe}). To show this, we make the standard approximation that the fraction of doubly-ionized oxygen can be estimated as
\begin{equation}
X(\rm{O}^{+2})\approx {\rm \frac{O^{+2}/H^+}{O^+/H^++O^{+2}/H^+}}\,,
\label{xo}
\end{equation}
i.e., we neglect the contributions by O$^0$ and O$^{3+}$ to the total oxygen abundance. These contributions are expected to be significant only in the cases of, respectively, very low and very high ionization parameter  \citep[e.g.][]{kobulnicky99,izotov06}. We adopt the prescription of \citet[][their equations~3 and 5]{izotov06} to compute the abundances of O$^+$ and O$^{+2}$ in equation~\eqref{xo} from observed emission-line luminosities, i.e.,
\begin{align}
\log {\rm O^+/H^+}= & \log \frac{L(\oiit)}{L(\hb)}-6.039+\frac{1.676}{\te} -0.40\log \te
\nonumber\\
                  & -0.034\te+\log (1+1.35x)
\label{oii}
\end{align}
and
\begin{align}
\log {\rm O^{+2}/H^+}= &\log \frac{L(\oiiitopt)}{L(\hb)}-5.800+\frac{1.251}{\te}
\nonumber\\
                  & -0.55\log \te-0.014\te\,,
\label{oiii}
\end{align}
where \te\ and $x$ have the same meaning as before. Fig.~\ref{fig:icf_obs_param}b shows the fraction of doubly-ionized oxygen estimated in this way, $X(\rm{O}^{+2})_{\rm Izotov}$, as a function of the true $X(\rm{O}^{+2})$, for the same models as in Fig.~\ref{fig:icf_obs_param}a. Again, as in Fig.~\ref{fig:c3o3_all} above, there is a large metallicity-dependent bias in the ionic abundance estimated using standard formulae and the direct-\Te\ method relative to the true one, except at solar-to-slightly sub-solar metallicity. 

Overall, therefore, Figs~\ref{fig:c3o3_all} and \ref{fig:icf_obs_param} indicate that estimates of the \COgas\ ratio via standard recipes involving the direct-\Te\ method are subject to strong biases and uncertainties affecting the derivation of both the \CppOpp\ ratio and the ICF in equation~\eqref{eq:co_from_icfr}, especially at non-solar metallicities. A full model of the type presented in this paper is required for more reliable abundance estimates at all metallicities. Interestingly, such a model offers the possibility to also explore in a physically consistent way the dependence of the ICF on the properties of the photoionized gas and the ionizing stellar populations.

\begin{figure}
\begin{center}
  \includegraphics[scale=0.43]{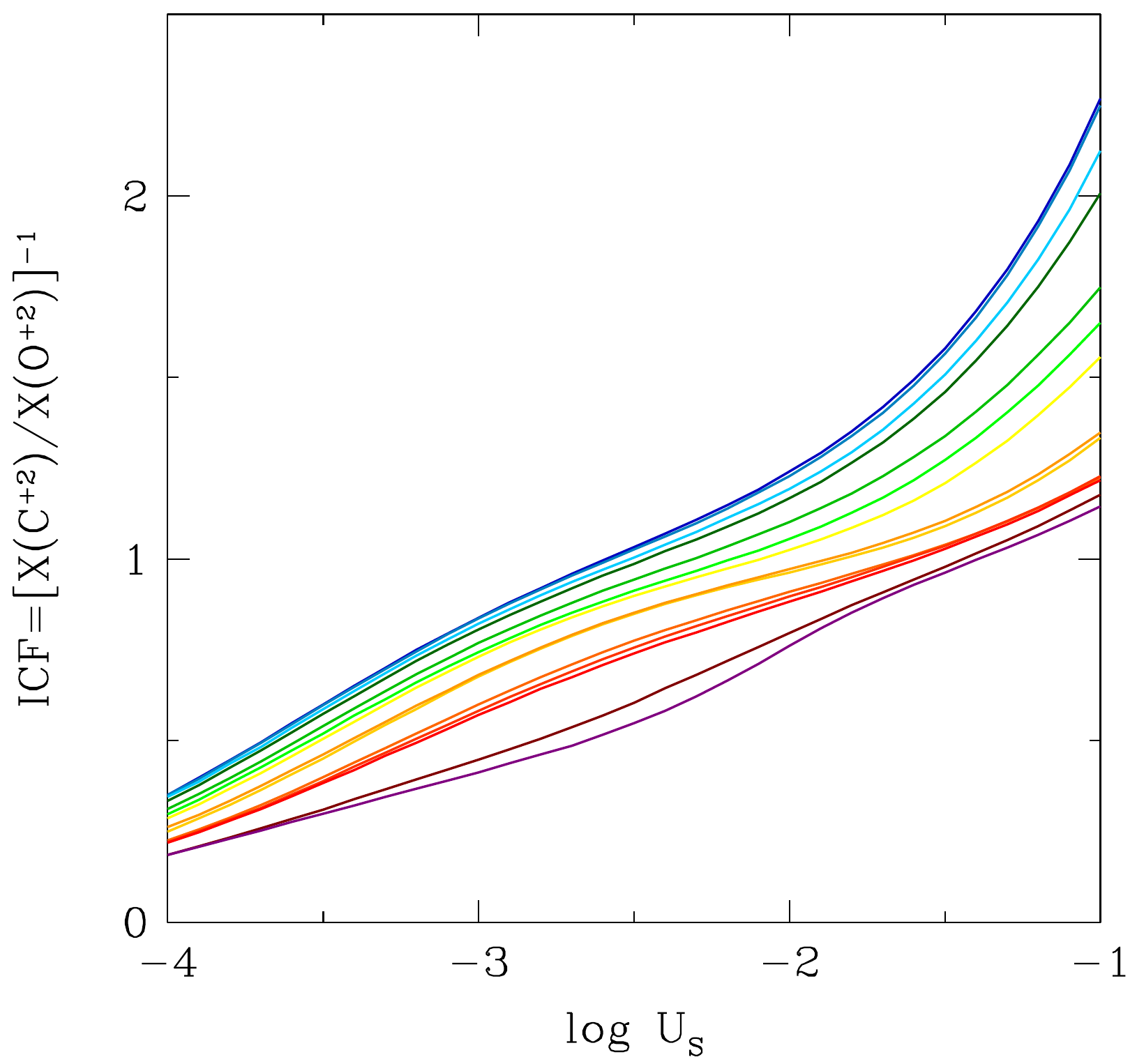}
  \caption[effect3]{Dependence of the ionization correction factor entering equation~\eqref{eq:co_from_icfr}, $[X(\rm{C}^{+2})/X(\rm{O}^{+2})]^{-1}$, on ionization parameter, $\log\Us$, for the full range of interstellar metallicities in Table~\ref{tab:parameters}, $0.0001\leq\zism\leq0.040$ (colour-coded as in Fig.~\ref{fig:NO_OH}), and for fixed dust-to-metal mass ratio, $\xid=0.3$, hydrogen density, $\nh=100\,{\rm cm}^{-3}$, carbon-to-oxygen ratio, $\COsol=0.44$, and IMF upper mass cutoff, $\mup=100\,\msol$.}
  \label{fig:icf_logU_allZ}
\end{center}
\end{figure}

\begin{figure}
\begin{center}
  \includegraphics[scale=0.43]{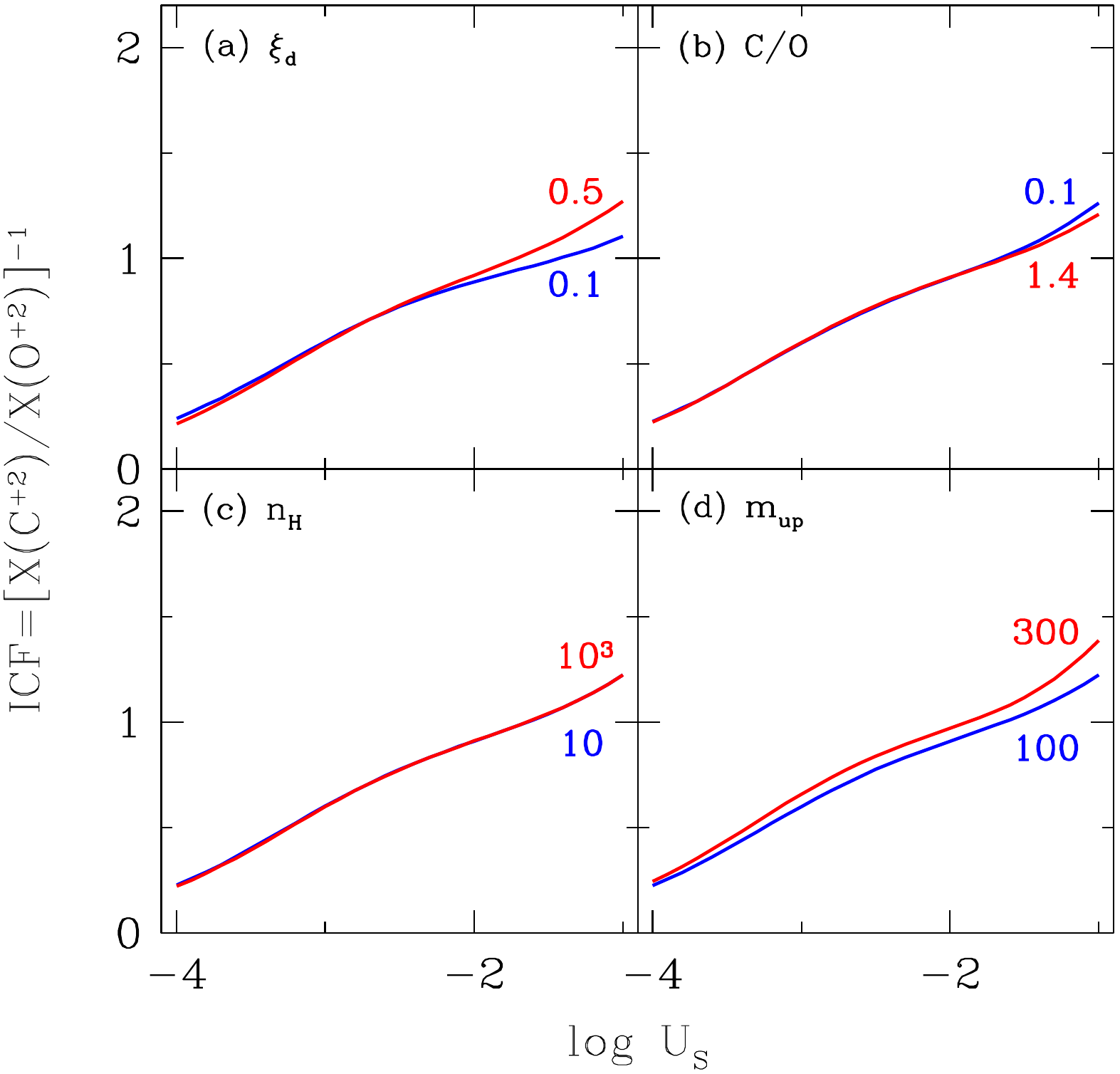}
  \caption{Dependence of the ionization correction factor entering equation~\eqref{eq:co_from_icfr}, $[X(\rm{C}^{+2})/X(\rm{O}^{+2})]^{-1}$, on ionization parameter, $\log\Us$, for the interstellar metallicity $\zism=0.014$ and for: (a) $\nh=100\,{\rm cm}^{-3}$, $\COsol=0.44$, $\mup=100\,\msol$ and two different dust-to-metal mass ratios, $\xid=0.1$ and 0.5; (b) $\xid=0.3$, $\nh=100\,{\rm cm}^{-3}$, $\mup=100\,\msol$ and two different carbon-to-oxygen ratios, $\CO=0.1$ and 1.4 times \COsol; (c) $\xid=0.3$, $\COsol=0.44$, $\mup=100\,\msol$ and two different hydrogen densities, $\nh=10$ and $10^3\,{\rm cm}^{-3}$; (d) $\xid=0.3$, $\nh=100\,{\rm cm}^{-3}$, $\COsol=0.44$, and two different IMF upper mass cutoffs, $\mup=100$ and $300\,\msol$}.
  \label{fig:icf_logU_param}
\end{center}
\end{figure}

To this end, we show in Fig.~\ref{fig:icf_logU_allZ} the dependence of the ICF on ionization parameter for models in a full range of interstellar metallicities, $0.0001\leq\zism\leq0.040$, and fixed dust-to-metal mass ratio, $\xid=0.3$, carbon-to-oxygen ratio, $\CO=\COsol$, hydrogen density, $\nh=100\,{\rm cm}^{-3}$, and IMF upper mass cutoff, $\mup=100\,\msol$. The trend of increasing ICF at increasing $\log\Us$ and decreasing \zism\ is the same as that described in the context of Fig.~\ref{fig:icf_obs_param} above. The added interest of Fig.~\ref{fig:icf_logU_allZ} is to explicit the strong dependence of the ICF on metallicity \citep[see also fig.~11 of][]{erb10}. In contrast, as Fig.~\ref{fig:icf_logU_param} shows, the other main adjustable parameter of the model have only a weak influence on the ICF, at fixed metallicity $\zism=0.014$. In particular, increasing \xid\ makes the gas-phase metallicity lower at fixed \zism, which leads to a slightly higher ICF, especially at large $\log\Us$ (see above; Fig.~\ref{fig:icf_logU_param}a). Also, since C is more depleted onto dust grains than O (Table~\ref{tab:abund_depl}), a drop in \CO\ ratio at fixed \zism\ and \xid\ causes the gas-phase metallicity to decrease a little, and hence, the ICF to rise (Fig.~\ref{fig:icf_logU_param}b). Changes in the hydrogen gas density have a negligible influence on the ICF (Fig.~\ref{fig:icf_logU_param}b). In contrast, raising the upper mass limit of the IMF from 100 to 300\,\msol\ makes the ionizing spectrum harder (Section~\ref{sec:optical_diagno}), causing more ionization of C to C$^{+3}$ and hence a greater ICF (Fig.~\ref{fig:icf_logU_param}d).

Finally, we stress that one of the unique features of the models presented in this paper regarding abundance estimates in star-forming galaxies is the possibility to relate emission-line measurements to not only gas-phase, but also interstellar (i.e. gas+dust-phase) element abundances. This is enabled by the careful parametrization of abundances and depletion factors described in Section~\ref{sec:abund} (Table~\ref{tab:abund_depl}). In the case of the \CO\ ratio, for example, since carbon is more depleted onto dust grains than oxygen, we find from the model library in Table~\ref{tab:parameters} that the interstellar \CO\ ratio is 1.06, 1.24 and 1.40 times larger than the gas-phase one, \COgas, for dust-to-metal mass ratios $\xid=0.1$, 0.3 and 0.5, respectively, and for $\zism=0.014$ and $\CO=\COsol$. These numbers change by only about 1 per cent across the whole metallicity range ($0.0001\leq\zism\leq0.040$; because of our inclusion of secondary nitrogen production) and for interstellar \CO\ ratios across the full range considered (between 0.1 and 1.4 times solar). 


\section{Conclusions}
\label{sec:concl}

We have presented a new model of the ultraviolet and optical nebular emission from star-forming galaxies, based on a combination of state-of-the-art stellar population synthesis and photoionization codes to describe the \hii\ regions and the diffuse gas ionized by successive stellar generations (following the approach of CL01). A main feature of this model is the self-consistent yet versatile treatment of element abundances and depletion onto dust grains, which allows one to relate the observed nebular emission from a galaxy to both gas-phase and dust-phase metal enrichment, over a wide range of chemical compositions. This feature should be particularly useful to investigate the early chemical evolution of galaxies with non-solar carbon-to-oxygen abundance ratios (e.g., \citealt{erb10,cooke11}; see also \citealt{garnett99}). In our model, the main adjustable parameters pertaining to the stellar ionizing radiation are the stellar metallicity, \zav, the IMF (the upper mass cutoff can reach $\mup=300\,\msol$) and the star formation history. The main adjustable  parameters pertaining to the ISM are the interstellar metallicity, \zism\ (taken to be the same as that of the ionizing stars), the zero-age ionization parameter of a newly born \hii\ region, \Us, the dust-to-metal mass ratio, \xid, the carbon-to-oxygen abundance ratio, \CO, and the hydrogen gas density, \nh. These should be regarded as `effective' parameters describing the global conditions of the gas ionized by young stars throughout the galaxy. 

We have built a comprehensive grid of photoionization models of star-forming galaxies spanning a wide range of physical parameters (Table~\ref{tab:parameters}). These models reproduce well the optical (e.g. \oiid, \hb, \oiii, \ha, \nii, \siid) and ultraviolet (e.g. \nv, \civd, \heii, \oiiis, \ciiid, \siliiid) emission-line properties of observed galaxies at various cosmic epochs. We find that ultraviolet emission lines are more sensitive than optical ones to parameters such as the \CO\ ratio, the hydrogen gas density, the upper IMF cutoff and even the dust-to-metal mass ratio, \xid. This implies that spectroscopic studies of the redshifted rest-frame ultraviolet emission of galaxies out to the reionization epoch should provide valuable clues about the nature of the ionizing radiation and early chemical enrichment of the ISM. In fact, the model presented in this paper has already been combined with a model of the nebular emission from narrow-line emitting regions of active galaxies to identify ultraviolet line-ratio diagnostics of photoionization by an AGN versus star formation \citep{feltre16}. It has also been used successfully to constrain the ionizing radiation and ISM parameters of galaxies at redshifts $2\lesssim z\lesssim9$, based on observed ultraviolet emission-line properties \citep{stark14,stark15a,stark15b,stark16}.

It is worth mentioning that the stellar population synthesis code used to generate the ionizing radiation in this paper (Section~\ref{sec:stellar_code}) does not incorporate binary stars, while models including binary stars \citep[e.g.,][]{eldridge08,eldridge12} have been preferred over existing `single-star' models to account for the observed rest-frame far-ultraviolet and optical composite spectra of a sample of 30 star-forming galaxies at redshift around $z=2.40$ \citep{steidel16}. We find that our new model can account remarkably well for the observed ultraviolet and optical emission-line properties of the composite \citet{steidel16} spectrum, with best-fitting metallicity $\zism=0.006$ (i.e. $0.4\zavsol$), ionization parameter $\log\Us=-3.0$ and carbon-to-oxygen ratio $\CO=0.52\COsol$.\footnote{Specifically, the predicted emission-line ratios of the best-fitting model are $\oiii/\hb=4.14$ (to be compared with the observed $4.25\pm0.09$), $\nii/\ha=0.09$ ($0.10\pm0.01$), $\siid/\ha=0.18$ ($0.18\pm0.01$), $\oiit/\oiii=0.68$ ($0.66\pm0.01$), $\siliiit/\ciiit=0.16$ ($0.33\pm0.07$), $\ciiit/\oiiit=4.05$ ($4.22\pm0.53$) and $\oiiit/\heii=6.84$ ($4.47\pm2.69$).} We further find that the predictions of our model for the \heii\ emission luminosity in low-metallicity star-forming galaxies are comparable to those of \citet{schaerer03}. For example, for a \citet{chabrier03} IMF truncated at 1 and 100\,\msol\ (very similar to the \citealt{salpeter55} IMF with same lower and upper cutoffs used by \citealt{schaerer03}) and  for the metallicity $\zav=0.001$, we find line luminosities between $\rm 2.7\times10^{38}\,erg\,s^{-1}$ and $\rm 4.04\times10^{39}\,erg\,s^{-1}$ per unit star formation rate (depending on \Us\ and \xid), consistent with the value $\rm 8.39\times10^{38}\,erg\,s^{-1}$ in table~4 of \citet[][his IMF `A']{schaerer03}. The calculations of \citet{schaerer03} extend all the way down to zero metallicity. For the smallest metallicity investigated here, $\zav=0.0001$, our model spans a range of \heii\ luminosities between $\rm 1.03\times10^{39}\,erg\,s^{-1}$ and $\rm 9.70\times10^{39}\,erg\,s^{-1}$, which can be compared to the values $\rm 2.91\times10^{37}$, $\rm 1.40\times10^{39}$ and $\rm 1.74\times10^{40}\,erg\,s^{-1}$ at $\zav=10^{-5}$, $10^{-7}$ and $0.$ in table~4 of \citet{schaerer03}.

Our fully self-consistent modeling of the nebular emission and chemical composition of the ISM in star-forming galaxies provides a unique way to test the reliability of standard recipes based on the direct-\Te\ method to measure element abundances from emission-line luminosities \citep[e.g.,][]{aller84,garnett95icf,izotov99,izotov06,shapley03,erb10}. We find that, for gas-phase metallicities around solar to slightly sub-solar, widely used formulae to constrain oxygen ionic fractions and the \CO\ ratio from ultraviolet and optical emission-line luminosities are reasonably faithful. However, the recipes break down at non-solar metallicities (both low and high; see Section~\ref{sec:COicf}), making their application inappropriate to studies of chemically young galaxies. In such cases, a fully self-consistent model of the kind presented in this paper is required to interpret the observed nebular emission. This can be achieved in an optimal way by appealing to a dedicated spectral analysis tool, such as the \beagle\ tool of \citet{chevallard16}, which already incorporates our model. Finally, we note that, while all the calculations presented in this paper pertain to ionization-bounded galaxies, our model provides a unique means of investigating the spectral signatures of the escape of ionizing photons from density-bounded galaxies. This will be the subject of a forthcoming study.

The model grid presented in this paper is available electronically from \url{http://www.iap.fr/neogal/models.html}.

\section*{Acknowledgements}

We thank the anonymous referee for helpful comments. We also thank Jarle Brinchmann, Anna Feltre, Brent Groves, Alba Vidal-Garc\'{i}a, Aida Wofford and the entire NEOGAL team for valuable discussions. We acknowledge support from the ERC via an Advanced Grant under grant agreement no. 321323-NEOGAL. GB acknowledges the hospitality and support from CNRS/IAP during different stages of this investigation, as well as support for this work from the National Autonomous University of M\'exico (UNAM) through grant PAPIIT IG100115.

Funding for the creation and distribution of the SDSS Archive has been provided by the Alfred P. Sloan Foundation, the Participating Institutions,  the  National  Aeronautics  and  Space  Administration, the National  Science  Foundation, the US Department of Energy, the Japanese Monbukagakusho and the Max Planck Society. The SDSS website is \url{http://www.sdss.org/}.



\bsp
\label{lastpage}
\end{document}